\documentclass[useAMS,usenatbib]{mn2e}
\usepackage{epsf,psfrag,graphicx}
\usepackage[usenames]{color}

\newcommand{\ifm}[1]{\relax\ifmmode#1\else$\mathsurround=0pt #1$\fi}
\newcommand{\pc}{\,{\rm pc}}
\newcommand{\Mv}{\hbox{$M_{\rm vir}$}}
\newcommand{\kpc}{\,{\rm kpc}}

\newcommand{\prop}{\propto}

\def\gtrsim{\lower.5ex\hbox{$\; \buildrel > \over \sim \;$}}

\def\MN{MN}
\def\CD{ART}
\def\C14{A$_{\rm RP}$}
\def\cmmc{\,{\rm cm}^{-3}}
\def\msun{{\rm M_\odot}}

\title[Inflow velocities of cold streams]
{Inflow velocities of cold flows streaming into massive galaxies at high
redshifts}

\author[Tobias Goerdt \& Daniel Ceverino]
{\parbox[t]{\textwidth}{Tobias Goerdt$^1$\thanks{tobias.goerdt@univie.ac.at}
and Daniel Ceverino$^2$}\\
\vspace*{3pt} \\
$^1$Institut f\"ur Astrophysik, T\"urkenschanzstra{\ss}e 17, Universit\"at
Wien, 1180 Wien, \"Osterreich\\
$^2$Centro de Astrobiolog{\'i}a (CSIC-INTA), Ctra de Torrej{\'o}n a Ajalvir, km
4, 28850 Torrej{\'o}n de Ardoz, Madrid, Espa\~na\\}

\date{Draft version \today}
\pagerange{\pageref{firstpage}--\pageref{lastpage}}
\pubyear{2015}

\begin{document}

\maketitle

\label{firstpage}

\begin{abstract}
We study the velocities of the accretion along streams from the cosmic web into
massive galaxies at high redshift with the help of three different suites of
\textsc{amr} hydrodynamical cosmological simulations. The results are compared
to free-fall velocities and to the sound speeds of the hot ambient medium. The
sound speed of the hot ambient medium is calculated using two different methods
to determine the medium's temperature. We find that the simulated cold stream
velocities are in violent disagreement with the corresponding free-fall
profiles. The sound speed is a better albeit not always correct description of
the cold flows' velocity. Using these calculations as a first order
approximation for the gas inflow velocities $v_{\rm inflow} = 0.9 \ v_{\rm vir}$
is given. We conclude from the hydrodynamical simulations as our main result
that the velocity profiles for the cold streams are constant with radius. These
constant inflow velocities seem to have a ``parabola-like'' dependency on the
host halo mass in units of the virial velocity that peaks at $M_{\rm vir} =
10^{12}$ M$_\odot$ and we also propose that the best fit functional form for the
dependency of the inflow velocity on the redshift is a square root power law
relation: $v_{\rm inflow} \propto \sqrt{z + 1} \ v_{\rm vir}$.
\end{abstract}

\begin{keywords}
cosmology: theory -- galaxies: evolution -- galaxies: formation --
galaxies: high redshift -- intergalactic medium -- methods: numerical 
\end{keywords}

\section{Introduction}
\label{sec:intro}

In the recent years it has been shown in theoretical work and in simulations
\citep{fardal, bd03, keresa, keresb, db06, ocvirk, DekelA_09a, dekel13} that
massive galaxies in the high redshift regime $(z \gtrsim 1.5)$, acquire their
baryons primarily via cold narrow streams of relatively dense and pristine gas
with temperatures around $10^4$ K. These cold flows penetrate through the
diffuse shock-heated medium. Their activity peaks around redshift 3. Having
reached the inner parts of the host halo their gas will eventually form a
dense, unstable, turbulent disc with a bulge where rapid star formation is
triggered \citep{oscara, oscarb, DekelA_09b, cd, andi, c12, cip, c14b,
marcello, genel12, genzel11, mandelker}. 

Galaxies' gas depletion time scales are understood to be short with respect to
the Hubble time throughout cosmic history \citep{daddi, genzel10}. So it
becomes apparent that galaxies must accrete fresh gas from the intergalactic
medium to sustain the observed levels of star formation over such a long time,
since otherwise their gas reservoirs would have been completely emptied. In the
cold flow scenario star formation behaves like a shower that is not draining
properly: no matter what are the initial conditions, a steady state will be
reached at the point where the amount of material that comes in via cold
streams is equal to the drained material, i.e. the gas that is being lost by
star formation and outflows. In the bathtub it is that the pressure at the
bottom that rises as it gets more full, pushing out more water. For galaxies,
it is the star formation rate that increases with the increase of available
gas. The result is a self-regulated system, the accretion is driving the
star-formation rate and not the amount of available gas \citep{bouche}. It is
in this sense that the star formation rate of a galaxy is fundamentally limited
by its gas inflow and does not depend so much on local physics, such as how
well the feedback regulates it. The timescales for star formation as well as
the star formation rate both depend crucially on the gas accretion rate, which
in turn depends on the speed of the infalling gas when it reaches the halo
centre. So doubling the infall velocity would also result in doubling the star
formation rate assuming constant cold stream cross sections and densities,
which highlights the importance of the inflow velocity.

Processes of observing the cold accretion stream paradigm in the real universe
are ongoing. \citet{mich2} made theoretical predictions about the likelihood of
observing these streams in absorption. Striking features in absorption have
indeed been observed \citep{bouche2}. \citet{ibata} found the existence of a
planar subgroup of satellites in the Andromeda galaxy (M 31), comprising about
half the population. \citet{mich3} demonstrated that this vast thin disk of
satellites can naturally be explained within the cold stream framework and
should therefore be interpreted as indirect observational evidence for the cold
stream paradigm.

\begin{figure*}
\begin{center}
\includegraphics[width=17.73cm]{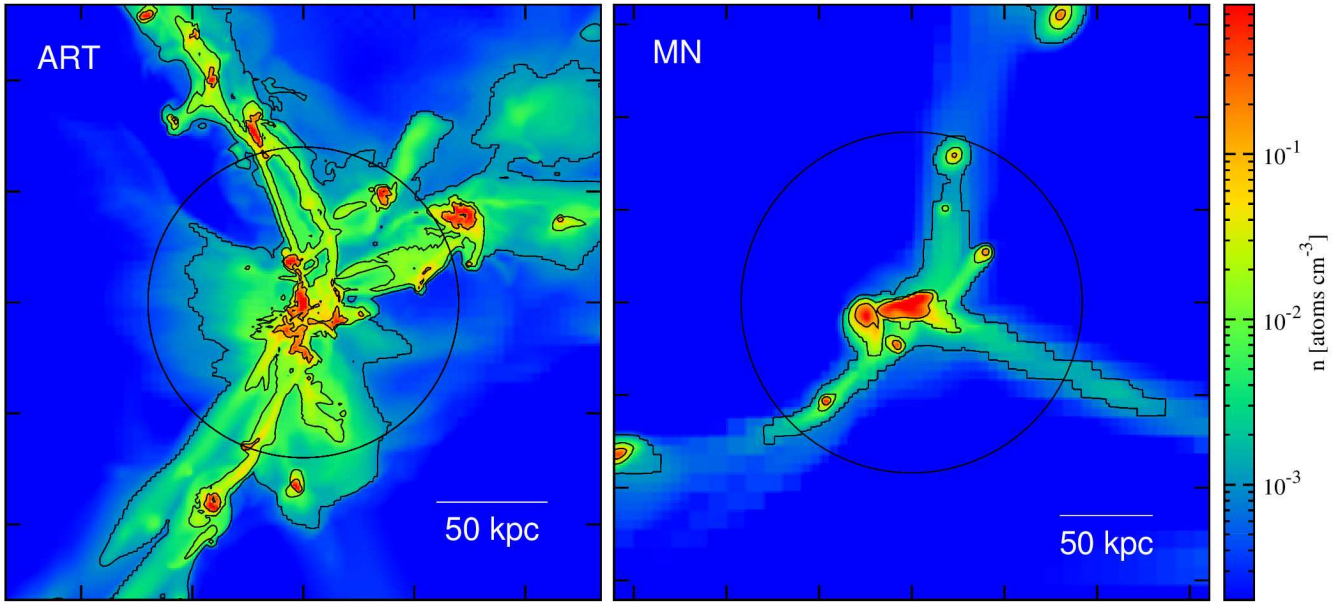}
\end{center}
\caption{Two simulated galaxies from {\CD} and {\MN}. The colour refers to the
maximum gas density along the line of sight. Contour lines are at $n = 0.1$,
0.01 and $0.001\cmmc$, respectively. The circles show the virial radii. Left: a
typical {\CD} galaxy (resolution $70\pc$) at $z=2.3$, with $\Mv = 3.5 \times
10^{11}\msun$. Right: one of the {\MN} galaxies (resolution $1\kpc$) at $z =
2.5$, with $\Mv = 10^{12}\msun$. In both cases, the inflow is dominated by
three cold narrow streams that are partly clumpy. The density in the streams is
$n = 0.003-0.1 \cmmc$, with the clump cores reaching $n \sim 1\cmmc$.}
\label{fig:denmap}
\end{figure*}

\citet{mich} used the cold stream model to explain the observational phenomenon
of Ly$\alpha$-blobs (LABs). They deployed cosmological hydrodynamical
\textsc{amr} simulations to predict the characteristics of Ly$\alpha$ emission
from the cold gas streams. These authors pointed out that in their simulations
the velocities of the cold streams are constant. So the potential energy of the
inflowing gas is not converted into kinetic energy. Subsequently they concluded
that it is the release of gravitational energy as the gas is flowing down the
potential gradient towards the halo centre which powers the Ly$\alpha$
luminosity. The potential energy of the inflowing gas is released as ly$\alpha$
radiation. This model was put forward as an alternative to other models trying
to explain the emission of LABs like being produced in the galaxy's
H\,\textsc{ii} regions but then being scattered in our direction by the
H\,\textsc{i} gas in the galaxy's circum-galactic medium \citep{steidel}, being
lit up due to illumination from a nearby buried AGN \citep{basu, scarlata,
prescott} or being lit up due to stellar feedback \citep{chapman, geacha,
geachb}.

\citet{faucher} tested the results of \citet{mich}. They analysed
cosmological smoothed particle hydrodynamics simulations to predict the
Ly$\alpha$ cooling emission of forming galaxies with the help of a Ly$\alpha$
radiative transfer code. In contrast to \citet{mich} they mention that a
scenario in which the cold streams free-fall into the haloes and release little
energy before hitting the central disk in a strong shock, may be supported by
their results. Later also \citet{joki} addressed the same question
independently. For this purpose they ran and analysed cosmological
radiation-hydrodynamics simulations. They looked at phase diagrams for their
haloes of gas speed versus radius in which they were able to distinguish
several cold streams. They produced corresponding free-fall profiles (i.e. the
speeds that the streams would follow if they were in free-fall). In contrast to
\citet{mich} they concluded that the streams are close to free-fall. As we will
also discuss in greater depth in section \ref{sec:inflvel}, the discrepancies
between their work and the results presented later in this paper might result
from the fact, that \citet{faucher} on the one hand deployed \textsc{sph}
simulations which used to have inaccuracies modelling dissipative processes
\citep{suckiness} and the sample of galaxies of \citet{joki} on the other hand
was very small (only three galaxies in total) and they report that not all of
their galaxies showed signs of free-fall behaviour.

Radial velocities around 10$^{12}$ M$_\odot$ haloes have been studied
\citep{freke}. They found roughly constant values of the radial inflow
velocities, with a decrease towards smaller radii.

\citet{nelsona} compare results from \textsc{sph} with moving mesh simulations
and find that the filamentary geometry of accreting gas near the virial radius
is a common feature in massive haloes above 10$^{11.5}$ M$_\odot$. Gas filaments
in \textsc{gadget}, however, tend to remain collimated and flow coherently to
small radii, or artificially fragment and form a large number of purely
numerical ‘blobs’. These same filamentary gas streams in \textsc{arepo} show
increased heating and disruption at 0.25 - 0.5 $r_{\rm vir}$ and contribute to
the hot gas accretion rate in a manner distinct from classical cooling flows.

\citet{nelsonb} investigated how the way galaxies acquire their gas across
cosmic time in cosmological hydrodynamic simulations is modified by a
comprehensive physical model for baryonic feedback processes finding that
feedback strongly suppresses the raw, as well as the net, inflow of this
‘smooth mode’ gas at all redshifts, regardless of the temperature history of
newly acquired gas.

\citet{wetzel} explored systematically the physics that governs cosmic
accretion into haloes and their galaxies. They analysed a suite of cosmological
simulations that incorporate both dark matter and gas dynamics with differing
treatments of gas cooling, star formation and thermal feedback to explore the
physics that governs the accretion of dark matter and baryons into halos and
their galaxies. They presented average radial velocities of the inflowing
material (their figures 4 and 6, bottom panels). The inflow velocities
presented there decrease with decreasing radius (roughly a decrease of 65\%
between 1 $r_{\rm vir}$ and the centre of the host).

In a companion paper \citep{mich5} we look at the amount of inflow -- the mass
accretion rate -- both as a function of radius, mass and redshift for the three
constituents gas, stars and dark matter.

In this paper we look at the velocity profiles of the gas inflow, both as a
function of radius, mass and redshift for the three constituents of the
simulations, namely gas, stars and dark matter. This study is worthwhile doing
since multiphase gas accretion is an interesting topic in galaxy formation
theory by itself and this study in particular has important implications on the
judgement of plausibility of both star formation rates as well as the
gravitational cooling model for LABs, as already mentioned. The paper is
organised as follows: in section \ref{sec:sim} we present the three suites of
simulations used for the following analysis. In section \ref{sec:mar} we show a
brief analysis of the mass accretion rate of the gas. In section \ref{sec:toym}
we present our toy models. In section\ref{sec:inflvel} we discuss the velocity
profiles of the inflow along the streams and in section \ref{sec:conc} we draw
our conclusions.

\section{simulations}
\label{sec:sim}

We use snapshots of galaxies from three different sets of simulations, all
three employing Eulerian \textsc{amr} (Adaptive Mesh Refinement) hydrodynamics
in a cosmological setting. The {\CD} \citep{cd, c12, dekel13, c14b} suite
consists of several zoom-in simulations with a maximum resolution of $35-70\pc$
at $z = 2$. The simulation zooms in on individual galaxies that reside in
dark-matter haloes which have masses of $(0.13 - 1.60) \times 10^{12}\msun$ at
$z = 2.3$. This {\CD} suite has got an extension that we will also use, the
{\C14} suite of simulations which includes radiation pressure feedback
\citep{cip}. This second suites consist of individual galaxies that reside in
dark-matter haloes of masses $(1 - 8) \times 10^{11}\msun$ at $z = 2.0$. The
third simulation is the Horizon-MareNostrum (hereafter \MN) simulation
\citep{ocvirk} containing dozens of massive galaxies in a cosmological box of
side 71 Mpc with a maximum resolution of $\le 1\kpc$ .

Density maps of galaxies of the {\MN} as well as the {\CD} suite are shown in
figure \ref{fig:denmap}. They demonstrate the dominance of typically three,
narrow cold streams, which come from well outside the virial radius along the
dark-matter filaments of the cosmic web, and penetrate into the discs at the
halo centres. The streams are partly clumpy and partly smooth, even in the
simulation of higher resolution. The typical densities in the streams are in
the range $n = 0.01 - 0.1\cmmc$, and they reach $n=0.1-1\cmmc$ at the clump
centres and in the central disk.

\subsection{High-resolution {\CD} simulations}

These simulations were run with the \textsc{amr} code \textsc{art}
\citep[Adaptive Refinement Tree;][]{kkk,andrey} with a spatial resolution better
than $70 \pc$ in physical units. It incorporates gas cooling, photoionisation
heating, star formation, metal enrichment and stellar feedback \citep{cak}
which are relevant physical processes for galaxy formation. For the gas
density, temperature, metallicity, and UV background in any given cell cooling
rates were computed using \textsc{cloudy} \citep{ferland}. It is assumed that
cooling takes place at the centre of a cloud of thickness 1 kpc
\citep{ceverino, rk}. Metallicity dependent, metal-line cooling is also
included, assuming a relative abundance of elements equal to the solar
composition. The code implements a ``constant" feedback model, in which the
combined energy from stellar winds and supernova explosions is released as a
constant heating rate over 40 Myr (the typical age of the lightest star that
can still explode in a type-II supernova). Photo-heating is taken into account
self-consistently with radiative cooling. A uniform UV background
\citep{haardt} is assumed, ignoring local sources. The self-shielding of dense,
galactic neutral hydrogen from the cosmological UV background
is mimicked by assuming a substantially suppressed UV background ($5.9 \times
10^{26} \ {\rm erg} \ {\rm s}^{-1} \ {\rm cm}^{-2} \ {\rm Hz}^{-1}$, the value of
the pre-reionisation UV background at $z=8$, according to the Haard \& Madau
model) for the gas at densities above $n = 0.1$ cm$^{-3}$. This density
threshold is motivated by radiative transfer calculations \citep[appendix A1
of][figure A2]{fumagalli}.

Our special version of the \textsc{art} code has a unique feature for the
purpose of simulating the detailed structure of the streams. It allows gas
cooling to well below $10^4$K so enabling the formation of high densities in
pressure equilibrium with the hotter and more dilute medium. A non-thermal
pressure floor has been implemented to ensure that the Jeans length is resolved
by at least seven resolution elements and thus prevent artificial fragmentation
on the smallest grid scale \citep{truelove,rk,cd}. It is effective in the dense
$(n > 10$ cm$^{-3})$ and cold $(T<10^4 K)$ regions inside galactic disks.

The equation of state remains unchanged at all densities. Stars form according
to a stochastic model that is described in appendix A of \citet{cak}. The SFR
surface density and the gas surface density are measured from the simulations
and compared with the Kennicutt relation in the appendix A of \citet{cip},
figure A1. The stochastic star formation model reproduces the \citet{kennicutt}
law. This happens in cells where the gas temperature is below $10^4$K and the
gas density is above a threshold of $n = 1\cmmc$. We use the IMF from
\citet{miller} that is also matching the results of \citet{woosley}. Metals
from supernovae type II and type Ia enrich the ISM. They are released from each
star particle by SNII at a constant rate for 40 Myr after its birth. The metal
ejection by SNIa assumes an exponentially declining SNIa rate from a maximum at
1 Gyr. The code treats the advection of metals self-consistently and it
distinguishes between SNII and SNIa ejecta \citep{ceverino}.

\subsubsection{Simulations including radiation pressure (\C14)}

The {\C14} suite of simulations \citep{cip} is a further development of last
subsection's {\CD} suite: apart from the features already presented there, it
also includes the effects of radiation pressure by massive stars. The radiation
pressure was modelled as a non-thermal pressure that acts only in dense and
optically thick star-forming regions in a way that the ionising radiation
injects momentum around massive stars, pressurising star-forming regions
\citep[][their appendix B]{oscarc}. The adaptive comoving mesh has been refined
in the dense regions to cells of minimum size between 17-35 pc in physical
units. The DM particle mass is $8.3 \times 10^4$ M$_\odot$. The particles
representing star clusters have a minimum mass of $10^3$ M$_\odot$, similar to
the stellar mass of an Orion-like star cluster.

\subsection{Ramses Horizon-MareNostrum simulation}

\begin{table}
\begin{center}
\setlength{\arrayrulewidth}{0.5mm}
\begin{tabular}{rcccr}
\hline
label & suite & $M_{\rm vir}$ [10$^{12}$ M$_\odot$] & $z$ & $N_{\rm gal}$ \\
\hline
10$^{13}$            & {\MN} & 10.47 $\pm$ 0.56   & 1.57 & 12 \\
10$^{13}$            & {\MN} & 10.49 $\pm$ 0.93   & 2.46 & 12 \\
$5 \times 10^{12}$   & {\MN} & 5.00 $\pm$ 0.045 & 1.57 & 12 \\
$5 \times 10^{12}$   & {\MN} & 5.48 $\pm$ 0.26   & 2.46 & 11 \\
10$^{12}$            & {\MN} & 1.03 $\pm$ 0.003 & 1.57 &  8 \\
10$^{12}$            & {\MN} & 1.01 $\pm$ 0.004 & 2.46 & 12 \\
10$^{12}$            & {\MN} & 1.03 $\pm$ 0.006 & 4.01 &  9 \\
10$^{11}$            & {\MN} & 0.099 $\pm$ 0.000 & 1.57 & 12 \\
10$^{11}$            & {\MN} & 0.099 $\pm$ 0.000 & 2.46 &  7 \\
10$^{11}$            & {\MN} & 0.099 $\pm$ 0.000 & 4.01 & 12 \\
$1.9 \times 10^{12}$ & {\CD} & 1.907 $\pm$ 0.217 & 1.14 $\pm$ 0.02  & 34 \\
$1.3 \times 10^{12}$ & {\CD} & 1.286 $\pm$ 0.093 & 1.60 $\pm$ 0.02  & 73 \\
$8.6 \times 10^{11}$ & {\CD} & 0.863 $\pm$ 0.046 & 2.25 $\pm$ 0.02  & 109 \\
$3.9 \times 10^{11}$ & {\CD} & 0.391 $\pm$ 0.034 & 3.40 $\pm$ 0.04  & 119 \\
$7.1 \times 10^{11}$ & {\C14} & 0.707 $\pm$ 0.055 & 1.14 $\pm$ 0.02  & 41 \\
$6.7 \times 10^{11}$ & {\C14} & 0.672 $\pm$ 0.049 & 1.58 $\pm$ 0.02  & 47 \\
$4.9 \times 10^{11}$ & {\C14} & 0.491 $\pm$ 0.028 & 2.21 $\pm$ 0.03  & 57 \\
$2.9 \times 10^{11}$ & {\C14} & 0.290 $\pm$ 0.019 & 3.27 $\pm$ 0.05  & 62 \\
$2.6 \times 10^{11}$ & {\C14} & 0.260 $\pm$ 0.010 & 1.13 $\pm$ 0.02  & 40 \\
$2.4 \times 10^{11}$ & {\C14} & 0.242 $\pm$ 0.010 & 1.59 $\pm$ 0.02  & 50 \\
$1.6 \times 10^{11}$ & {\C14} & 0.160 $\pm$ 0.007 & 2.29 $\pm$ 0.03  & 60 \\
$7.3 \times 10^{10}$ & {\C14} & 0.073 $\pm$ 0.004 & 3.53 $\pm$ 0.05  & 66 \\
\hline
\end{tabular}
\end{center}
\caption{The different bins of galaxies which are used throughout this paper.
'Label' denotes the tag a bin is labelled with in any of the figures. It
usually is a mass very close to the ensemble's actual mean virial mass. 'Suite'
indicates the simulation set the bin consists of. $M_{\rm vir}$ gives the actual
mean virial mass of the ensemble together with its standard deviation. $z$ is
the mean redshift of the ensemble together with its standard deviation. We do
not combine galaxies from different {\MN} snapshots therefore the standard
deviation of the redshift of any of the {\MN} bins is always zero. $N_{\rm gal}$
is the number of galaxies in the respective bin.}
\label{tab:bins}
\end{table}

The \textsc{amr} code \textsc{ramses} \citep{teyssier} is used for this
simulation. Its spatial resolution in physical units is $\sim 1\,$kpc. We
included UV heating via a background model \citep{haardt}. Supernovae feedback
and metal enrichment are modelled in a simple way using the implementation
described in \citet{dubois}. Cooling rates are calculated assuming ionisation
equilibrium for H and He, for collisional- and photo-ionisation \citep{katz}.
Metal cooling is assumed to be proportional to the metallicity, relative to the
solar abundance \citep{grevesse}. It is calculated with the help of tabulated
\textsc{cloudy} rates. Unlike in the {\CD} or {\C14} simulations, no cooling
below $T<10^4$K is computed, and no self-shielding of the UV flux is assumed.

For high-density regions, we consider a polytropic equation of state with
$\gamma_0 = 5/3$ to model the complex, turbulent multi-phase structure of the
inter-stellar medium (ISM) \citep{yepes, sh} in a simplified form
\citep[see][]{joop, dubois}. The ISM is defined as gas with hydrogen density
greater than $n_{\rm H} = 0.1\cmmc$. This definition is one order of magnitude
lower than the one of the {\CD} and {\C14} simulations. Star formation only for
ISM gas has been included. It is modelled by spawning star particles at a rate
consistent with the \citet{kennicutt} law derived from local observations of
star forming galaxies. 

The \textsc{ramses} code implements a pressure floor in order to prevent
artificial fragmentation, by keeping the Jeans lengthscale, $\lambda_{\rm J}
\prop T n^{-2/3}$, larger than four grid-cell sides everywhere. In any case
where $n>0.1\cmmc$, a density dependent temperature floor was imposed. It
mimics the average thermal and turbulent pressure of the multiphase ISM
\citep{sh, dvs}. In our case, we allow the gas to heat up above this
temperature floor and cool back. The temperature floor follows a polytropic
equation of state with $T_{\rm floor}=T_0 (n/n_0)^{\gamma_0-1}$, where $T_0=10^4$ K
and $n_0=$ 0.1 atoms cm$^{-3}$. The resulting pressure floor is given by $P_{\rm
floor} = n_{\rm H} k_{\rm B} T_{\rm floor}$. For each stellar population, 10\% of
the mass is assumed to go through a supernovae type II event after 10 Myr. For
each supernova, 10\% of the ejected mass is assumed to be pure metals. The
energy and the metals produced by the supernova event are released in a single
impulse. The remaining 90\% of the ejected mass keep the metallicity of their
star at birth. Feedback through supernovae type Ia has not been considered.

\begin{figure*}
\begin{center}
\psfrag{$0.2$}[B][B][1][0]{$0.2$}
\psfrag{$0.4$}[B][B][1][0]{$0.4$}
\psfrag{$0.6$}[B][B][1][0]{$0.6$}
\psfrag{$0.8$}[B][B][1][0]{$0.8$}
\psfrag{$1$}[B][B][1][0]{$1$}
\psfrag{$1.2$}[B][B][1][0]{$1.2$}
\psfrag{$1.4$}[B][B][1][0]{$1.4$}
\psfrag{$1.6$}[B][B][1][0]{$1.6$}
\psfrag{$20$}[B][B][1][0]{$20$}
\psfrag{$30$}[B][B][1][0]{$30$}
\psfrag{$40$}[B][B][1][0]{$40$}
\psfrag{$50$}[B][B][1][0]{$50$}
\psfrag{$60$}[B][B][1][0]{$60$}
\psfrag{$70$}[B][B][1][0]{$70$}
\psfrag{$80$}[B][B][1][0]{$80$}
\psfrag{$90$}[B][B][1][0]{$90$}
\psfrag{$100$}[B][B][1][0]{$100$}
\psfrag{gas}[Br][Br][1][0]{gas}
\psfrag{r [rvir]}[B][B][1][0]{$r$ [$r_{\rm vir}$]}
\psfrag{Minflo [MO/yr]}[B][B][1][0]{$\dot M$ [M$_\odot$ yr$^{-1}$]}
\psfrag{ARP}[Br][Br][1][0]{\C14}
\psfrag{M = 7.3e10 MO}[Br][Br][1][0]{$M=7.3\times10^{10}$ M$_\odot$}
\psfrag{M = 1.6e11 MO}[Br][Br][1][0]{$M=1.6\times10^{11}$ M$_\odot$}
\psfrag{M = 2.4e11 MO}[Br][Br][1][0]{$M=2.4\times10^{11}$ M$_\odot$}
\psfrag{M = 2.6e11 MO}[Br][Br][1][0]{$M=2.6\times10^{11}$ M$_\odot$}
\psfrag{M = 2.9e11 MO}[Br][Br][1][0]{$M=2.9\times10^{11}$ M$_\odot$}
\psfrag{M = 4.9e11 MO}[Br][Br][1][0]{$M=4.9\times10^{11}$ M$_\odot$}
\psfrag{M = 6.7e11 MO}[Br][Br][1][0]{$M=6.7\times10^{11}$ M$_\odot$}
\psfrag{M = 7.1e11 MO}[Br][Br][1][0]{$M=7.1\times10^{11}$ M$_\odot$}
\psfrag{z = 3.53}[Br][Br][1][0]{$z=3.53$}
\psfrag{z = 3.27}[Br][Br][1][0]{$z=3.27$}
\psfrag{z = 2.29}[Br][Br][1][0]{$z=2.29$}
\psfrag{z = 2.21}[Br][Br][1][0]{$z=2.21$}
\psfrag{z = 1.59}[Br][Br][1][0]{$z=1.59$}
\psfrag{z = 1.58}[Br][Br][1][0]{$z=1.58$}
\psfrag{z = 1.14}[Br][Br][1][0]{$z=1.14$}
\psfrag{z = 1.13}[Br][Br][1][0]{$z=1.13$}
\includegraphics[width=17.73cm]{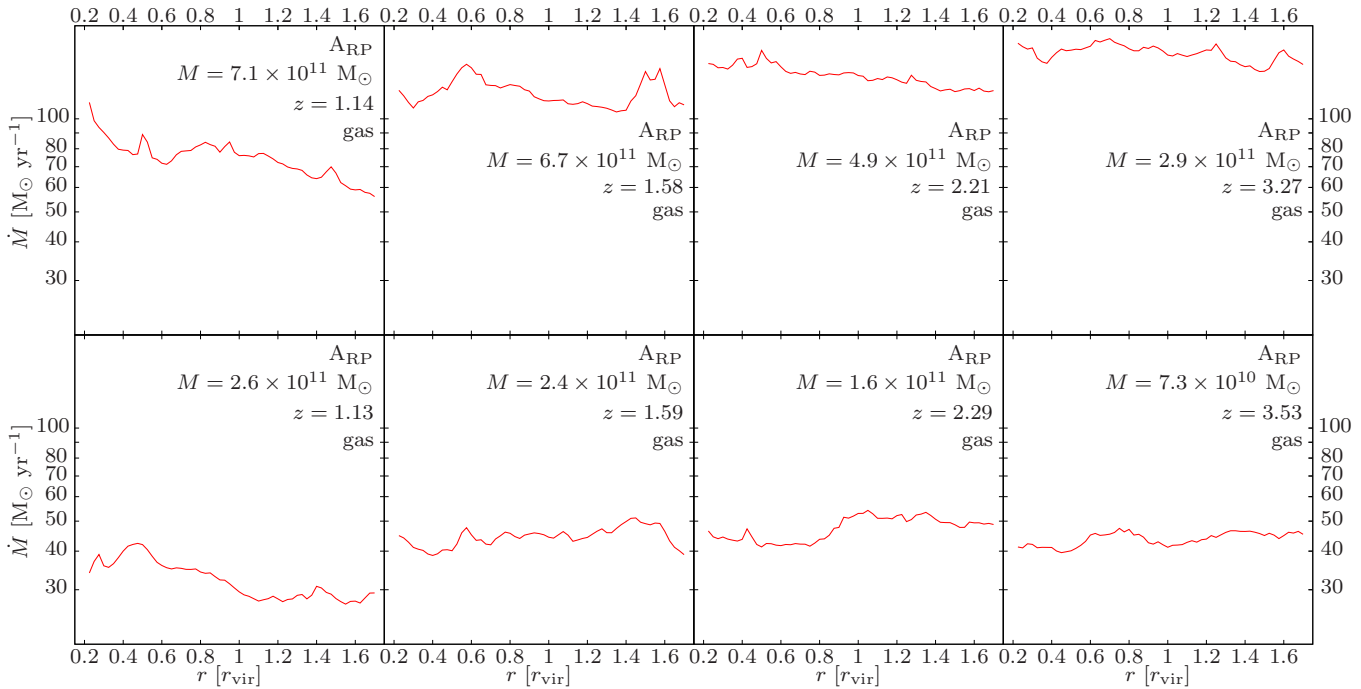}
\end{center}
\caption{Average inflow as a function of radius for different halo masses and
redshifts for the {\C14} simulations. The different panels indicate a variety
of different halo masses and redshifts. The inflow in gas is remarkably
constant with radius, except for the higher mass very low redshift panel, which
has a very mildly declining mass accretion with increasing radius. However by
and large the mass accretion rate of gas is constant with radius.}
\label{fig:avinflvsradARP}
\end{figure*}

The initial conditions were constructed assuming a $\Lambda$CDM universe with
$\Omega_{\rm M} = 0.3$, $\Omega_\Lambda = 0.7$, $\Omega_{\rm b} = 0.045$, $h=0.7$
and $\sigma_8 = 0.9$ in a periodic box of 71 Mpc. The adaptive-resolution rules
in this simulation were the same everywhere, with no zoom-in resimulation of
individual galaxies. The dark matter particle mass is $1.16 \times 10^7 \msun$,
the star particle mass is $2.05 \times 10^6 \msun$, the smallest cell size is
$1.09$ kpc physical, and the force softening length is 1.65 kpc.

We will show averaged results for an ensemble of galaxies having very similar
masses and redshifts. In table \ref{tab:bins} we show a summary of the various
bins of galaxies we use. Since we have better statistics for the {\MN} 
simulation we could bin galaxies from a narrower mass range therefore the
standard deviations of the mean mass of a {\MN} bin is usually much smaller. To
partly compensate for that we combine {\CD} as well as {\C14} galaxies from
adjacent redshifts increasing the statistics but introducing a standard
deviation into the mean redshift of the bin.

\section{Mass accretion rate}
\label{sec:mar}

In a companion paper \citep{mich5} we look at the amount of inflow -- the mass
accretion rate -- both as a function of radius, mass and redshift for the three
constituents gas, stars and dark matter. Since the mass accretion rate of the
gas is ultimately what is interesting for the rate of star formation and how
much gas “needs” to be lost via outflows we show briefly an analysis of the
mass accretion rate of the gas in this section.

The analysis of the {\C14} simulations starts by looking at the average gas
inflow as a function of radius: We measure the amount of mass which is crossing
a spherical shell of radius $r$ centred around a given individual galaxy within
a small time $\Delta t$. The mass crossing the shell is divided by the time
taken $(\Delta t)$ to get an inflow rate. In order to get the best estimate for
the inflow in case of gas we decided to take into account only the cells which
have an inwards radial velocity. In order to get rid of the statistical noise
of a single galaxies we average the amount of inflow over all available
galaxies having similar redshifts and masses.

The amount of inflow into the galaxies as a function of radius is shown in
figure \ref{fig:avinflvsradARP}. The magnitude of the inflow is basically
constant over all radii $r$ regardless of the actual mass-redshift bin, except
for the one higher mass very low redshift panel, which has a very mildly
declining mass accretion with increasing radius. However by and large the
amount of inflow is constant with radius. So we can mainly see that the mass
accretion rate simply follows the velocity profile, i.e. cold and hot gas are
not mixed as it falls in. As already mentioned we discuss these trends and also
the mass accretion rates of stars and dark matter in a companion paper
\citep{mich5}.

\section{Analytical models}
\label{sec:toym}

Naively one would expect the gas inflow velocities to obey one of the following
two quantities: either free-fall or the sound speed of the hot ambient medium.
The underlying assumption of the first case is that the gas flows freely within
the gravitational potential into its host and is otherwise completely
unconstrained. In the second case one assumes that the velocity of the gas will
be fundamentally limited by the sound speed of the hot ambient medium through
the effect of shocks which occur within the streams themselves as well as at
the boundary layers between the streams and the hot ambient medium as soon as
the velocity of the inflowing gas reaches the sound speed of this ambient
medium. 

\subsection{Free-fall}

Free-fall profiles cannot be calculated analytically. To evaluate them
numerically we follow the method laid out by \citet{joki} which we will quickly
outline here: a static, spherically symmetric potential is assumed. We
integrate the free-fall speeds $v_{\rm ff}(r)$ from a starting position $r_0$
and a starting velocity $v_{\rm init} = v_{\rm ff}(r_0)$ towards the halo centre
using
\begin{equation}
dv_{\rm ff}(r) = {1 \over v_{\rm ff}(r)}{G \ M(<r) \over r^2}dr,
\end{equation}
where $r$ is radius, $M (< r)$ is the total halo mass within $r$ and $G$ is the
gravitational constant. In practice we divide the host halo into spherical bins
at radial positions $r_{\rm i}$, where an increasing $i$ corresponds to a
decreasing radius, and solve the above equation by recursively computing
\begin{equation}
v_{\rm ff}(r_{\rm i + 1}) = v_{\rm ff}(r_{\rm i}) + {1 \over v(r_{\rm i})}{G \ M(<
r_{\rm i + 1}) \over r_{\rm i + 1}^2}(r_{\rm i + 1} - r_{\rm i}).
\end{equation}
The underlying mass distribution $M(< r_{\rm i + 1})$ is measured from the
outputs of the hydrodynamical simulations as described in section
\ref{sec:sim}. We choose an initial position $(r_0 = 1.7 \ r_{\rm vir})$ as well
as an initial inward velocity $(v_{\rm init} = 0)$ and integrate the path of the
free-falling gas iteratively until a very small radius $(0.1 \ r_{\rm vir})$ is
reached. The free-fall profiles usually show a strongly increasing velocity
with decreasing radius. The results of the calculations are presented in detail
in section \ref{sec:inflvel}. We will see that these predictions violently
disagree with the results from the hydrodynamical simulations.

\subsection{Sound speed of the ambient medium}

The sound speed $c_{\rm s}$ of the hot ambient halo gas on the other hand can be
calculated analytically. Sound speed as such equals to
\begin{equation}
c_{\rm s}\left(\vec{r}\right) = \sqrt{\gamma \ k_{\rm B} \ T(\vec{r}) \over
m_{\rm p} \ \mu_{\rm av}},
\label{eqn:cs}
\end{equation}
with $\gamma$ being the ratio of the heat capacity at constant pressure to the
heat capacity at constant volume, which is usually a constant $(\sim 5/3)$.
$k_{\rm B}$ is Boltzmann's constant, $m_{\rm p}$ is the proton mass, $\mu_{\rm av}$
is the mean molecular weight (for ionised gas $\mu_{\rm av} = 0.61)$ and
$T(\vec{r})$ is temperature of the gas as a function of position.

\subsubsection{Approximated constant halo gas temperature}

\begin{figure*}
\begin{center}
\psfrag{gas}[Br][Br][1][0]{\textcolor{red}{gas}}
\psfrag{star}[Br][Br][1][0]{\textcolor{green}{stars}}
\psfrag{DM}[Br][Br][1][0]{\textcolor{blue}{DM}}
\psfrag{free fall}[Br][Br][1][0]{\textcolor{magenta}{free-fall}}
\psfrag{cs hot gas}[Br][Br][1][0]{\textcolor{cyan}{$c_{\rm s}$ hot gas}}
\psfrag{$0.2$}[B][B][1][0]{$0.2$}
\psfrag{$0.4$}[B][B][1][0]{$0.4$}
\psfrag{$0.6$}[B][B][1][0]{$0.6$}
\psfrag{$0.8$}[B][B][1][0]{$0.8$}
\psfrag{$1$}[B][B][1][0]{$1$}
\psfrag{$1.2$}[B][B][1][0]{$1.2$}
\psfrag{$1.4$}[B][B][1][0]{$1.4$}
\psfrag{$1.6$}[B][B][1][0]{$1.6$}
\psfrag{r [rvir]}[B][B][1][0]{$r$ [$r_{\rm vir}$]}
\psfrag{vinflo [vvir]}[B][B][1][0]{$v_{\rm inflow}$ [$v_{\rm vir}$]}
\psfrag{MN}[Br][Br][1][0]{\MN}
\psfrag{M = 1e11 MO}[Br][Br][1][0]{$M = 10^{11}$ M$_\odot$}
\psfrag{M = 1e12 MO}[Br][Br][1][0]{$M = 10^{12}$ M$_\odot$}
\psfrag{M = 5e12 MO}[Br][Br][1][0]{$M = 5 \times 10^{12}$ M$_\odot$}
\psfrag{M = 1e13 MO}[Br][Br][1][0]{$M = 10^{13}$ M$_\odot$}
\psfrag{z = 1.57}[Br][Br][1][0]{$z = 1.57$}
\psfrag{z = 2.46}[Br][Br][1][0]{$z = 2.46$}
\psfrag{z = 4.01}[Br][Br][1][0]{$z = 4.01$}
\includegraphics[width=16.03cm]{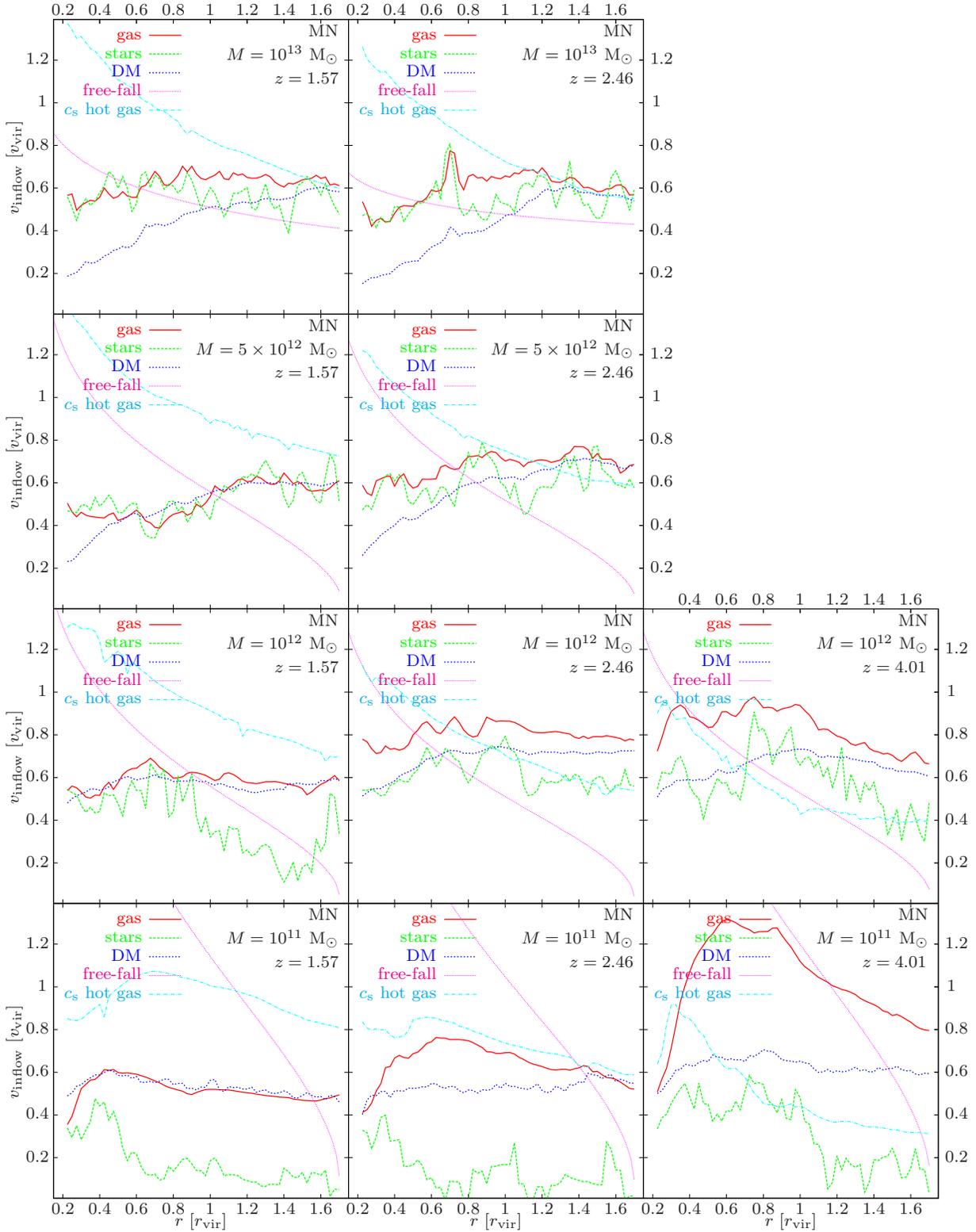}
\end{center}
\caption{Mass weighted averages of the inflow velocities as a function of
radius for the {\MN} simulations. The different panels indicate a variety of
different halo masses and redshifts. Corresponding free-fall profiles are
over-plotted in dotted magenta and profiles of the sound speed of the hot
ambient halo gas in dash-dotted turquoise. For low mass haloes $(M_{\rm vir} \le
10^{12}$ M$_\odot)$ at all redshifts the velocities are roughly constant with
radius. For high mass haloes $(M_{\rm vir} \ge 5 \times 10^{12}$ M$_\odot)$ the
inflow velocities increase slightly with increasing radius. Subsequently no
sign of free-fall can be seen in any of the bins. In the very high redshift and
low mass cases one sees an increase of velocity with decreasing radius down to
0.6 $r_{\rm vir}$ from where the velocity is decreasing again. These shapes are
typical for high redshift low mass haloes. In units of the virial velocity the
average inflow velocity stays roughly constant with varying host halo mass, it
increases with an increase in redshift.}
\label{fig:inflvelvsradMN}
\end{figure*}

\begin{figure*}
\begin{center}
\psfrag{$0.2$}[B][B][1][0]{$0.2$}
\psfrag{$0.4$}[B][B][1][0]{$0.4$}
\psfrag{$0.6$}[B][B][1][0]{$0.6$}
\psfrag{$0.8$}[B][B][1][0]{$0.8$}
\psfrag{$1$}[B][B][1][0]{$1$}
\psfrag{$1.2$}[B][B][1][0]{$1.2$}
\psfrag{$1.4$}[B][B][1][0]{$1.4$}
\psfrag{$1.6$}[B][B][1][0]{$1.6$}
\psfrag{gas}[Br][Br][1][0]{\textcolor{red}{gas}}
\psfrag{star}[Br][Br][1][0]{\textcolor{green}{stars}}
\psfrag{DM}[Br][Br][1][0]{\textcolor{blue}{DM}}
\psfrag{free fall}[Br][Br][1][0]{\textcolor{magenta}{free-fall}}
\psfrag{cs hot gas}[Br][Br][1][0]{\textcolor{cyan}{$c_{\rm s}$ hot gas}}
\psfrag{r [rvir]}[B][B][1][0]{$r$ [$r_{\rm vir}$]}
\psfrag{vinflo [vvir]}[B][B][1][0]{$v_{\rm inflow}$ [$v_{\rm vir}$]}
\psfrag{CDB}[Br][Br][1][0]{\CD}
\psfrag{M = 3.9e11 MO}[Br][Br][1][0]{$M=3.9\times10^{11}$ M$_\odot$}
\psfrag{M = 8.6e11 MO}[Br][Br][1][0]{$M=8.6\times10^{11}$ M$_\odot$}
\psfrag{M = 1.3e12 MO}[Br][Br][1][0]{$M=1.3\times10^{12}$ M$_\odot$}
\psfrag{M = 1.9e12 MO}[Br][Br][1][0]{$M=1.9\times10^{12}$ M$_\odot$}
\psfrag{z = 3.40}[Br][Br][1][0]{$z=3.40$}
\psfrag{z = 2.25}[Br][Br][1][0]{$z=2.25$}
\psfrag{z = 1.60}[Br][Br][1][0]{$z=1.60$}
\psfrag{z = 1.14}[Br][Br][1][0]{$z=1.14$}
\includegraphics[width=17.73cm]{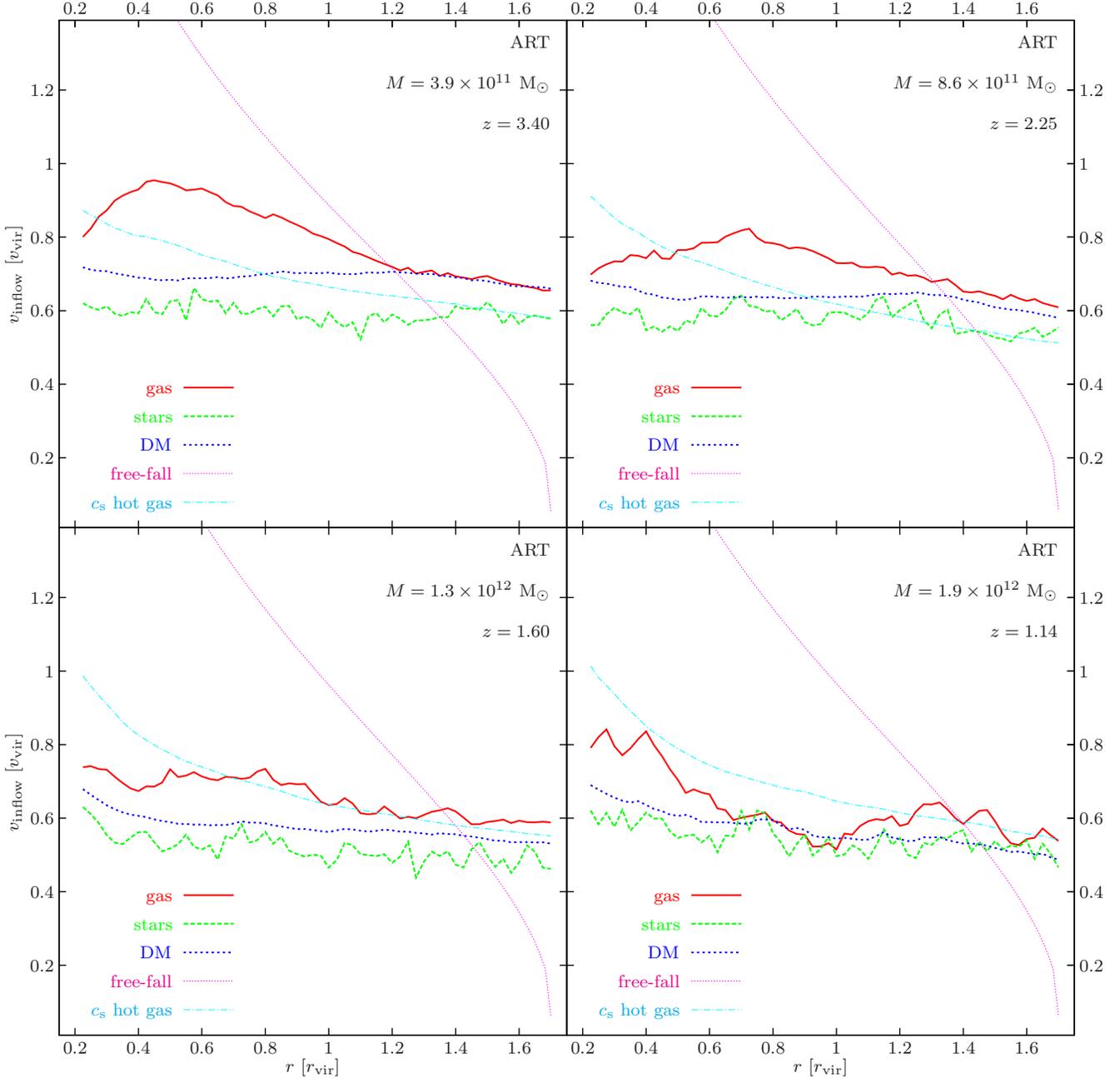}
\end{center}
\caption{Mass weighted averages of the inflow velocities as a function of
radius for the {\CD} simulations. Free-fall profiles are over-plotted in dotted
magenta and the sound speeds of the hot ambient halo gas in dash-dotted
turquoise. The inflow velocities of the gas increase slightly with decreasing
radius. The total increase over the whole radius range is usually roughly 50\%.
These increases however are not consistent with free-fall as the corresponding
profiles show. The sound speeds of the hot halo gas are a much better proxy, at
least at high redshift. In the higher redshift cases one sees an increase of
velocity with decreasing radius down to $\sim 0.5 \ r_{\rm vir}$ from where the
velocity is decreasing again. A shape like that is typical for low mass haloes
at high redshift, it originates in the loss of angular momentum during the
``strong-torque phase'' \citep{mark}.}
\label{fig:inflvelvsradCDB}
\end{figure*}

\begin{figure*}
\begin{center}
\psfrag{$0.2$}[B][B][1][0]{$0.2$}
\psfrag{$0.4$}[B][B][1][0]{$0.4$}
\psfrag{$0.6$}[B][B][1][0]{$0.6$}
\psfrag{$0.8$}[B][B][1][0]{$0.8$}
\psfrag{$1$}[B][B][1][0]{$1$}
\psfrag{$1.2$}[B][B][1][0]{$1.2$}
\psfrag{$1.4$}[B][B][1][0]{$1.4$}
\psfrag{$1.6$}[B][B][1][0]{$1.6$}
\psfrag{gas}[Br][Br][1][0]{\textcolor{red}{gas}}
\psfrag{star}[Br][Br][1][0]{\textcolor{green}{stars}}
\psfrag{DM}[Br][Br][1][0]{\textcolor{blue}{DM}}
\psfrag{free fall}[Br][Br][1][0]{\textcolor{magenta}{free-fall}}
\psfrag{cs hot gas}[Br][Br][1][0]{\textcolor{cyan}{$c_{\rm s}$ hot gas}}
\psfrag{r [rvir]}[B][B][1][0]{$r$ [$r_{\rm vir}$]}
\psfrag{vinflo [vvir]}[B][B][1][0]{$v_{\rm inflow}$ [$v_{\rm vir}$]}
\psfrag{ARP}[Br][Br][1][0]{\C14}
\psfrag{M = 7.1e11 MO}[Br][Br][1][0]{$M=7.1\times10^{11}$ M$_\odot$}
\psfrag{M = 6.7e11 MO}[Br][Br][1][0]{$M=6.7\times10^{11}$ M$_\odot$}
\psfrag{M = 4.9e11 MO}[Br][Br][1][0]{$M=4.9\times10^{11}$ M$_\odot$}
\psfrag{M = 2.9e11 MO}[Br][Br][1][0]{$M=2.9\times10^{11}$ M$_\odot$}
\psfrag{M = 2.6e11 MO}[Br][Br][1][0]{$M=2.6\times10^{11}$ M$_\odot$}
\psfrag{M = 2.4e11 MO}[Br][Br][1][0]{$M=2.4\times10^{11}$ M$_\odot$}
\psfrag{M = 1.6e11 MO}[Br][Br][1][0]{$M=1.6\times10^{11}$ M$_\odot$}
\psfrag{M = 7.3e10 MO}[Br][Br][1][0]{$M=7.3\times10^{10}$ M$_\odot$}
\psfrag{z = 3.53}[Br][Br][1][0]{$z=3.53$}
\psfrag{z = 3.27}[Br][Br][1][0]{$z=3.27$}
\psfrag{z = 2.29}[Br][Br][1][0]{$z=2.29$}
\psfrag{z = 2.21}[Br][Br][1][0]{$z=2.21$}
\psfrag{z = 1.59}[Br][Br][1][0]{$z=1.59$}
\psfrag{z = 1.58}[Br][Br][1][0]{$z=1.58$}
\psfrag{z = 1.14}[Br][Br][1][0]{$z=1.14$}
\psfrag{z = 1.13}[Br][Br][1][0]{$z=1.13$}
\includegraphics[width=17.73cm]{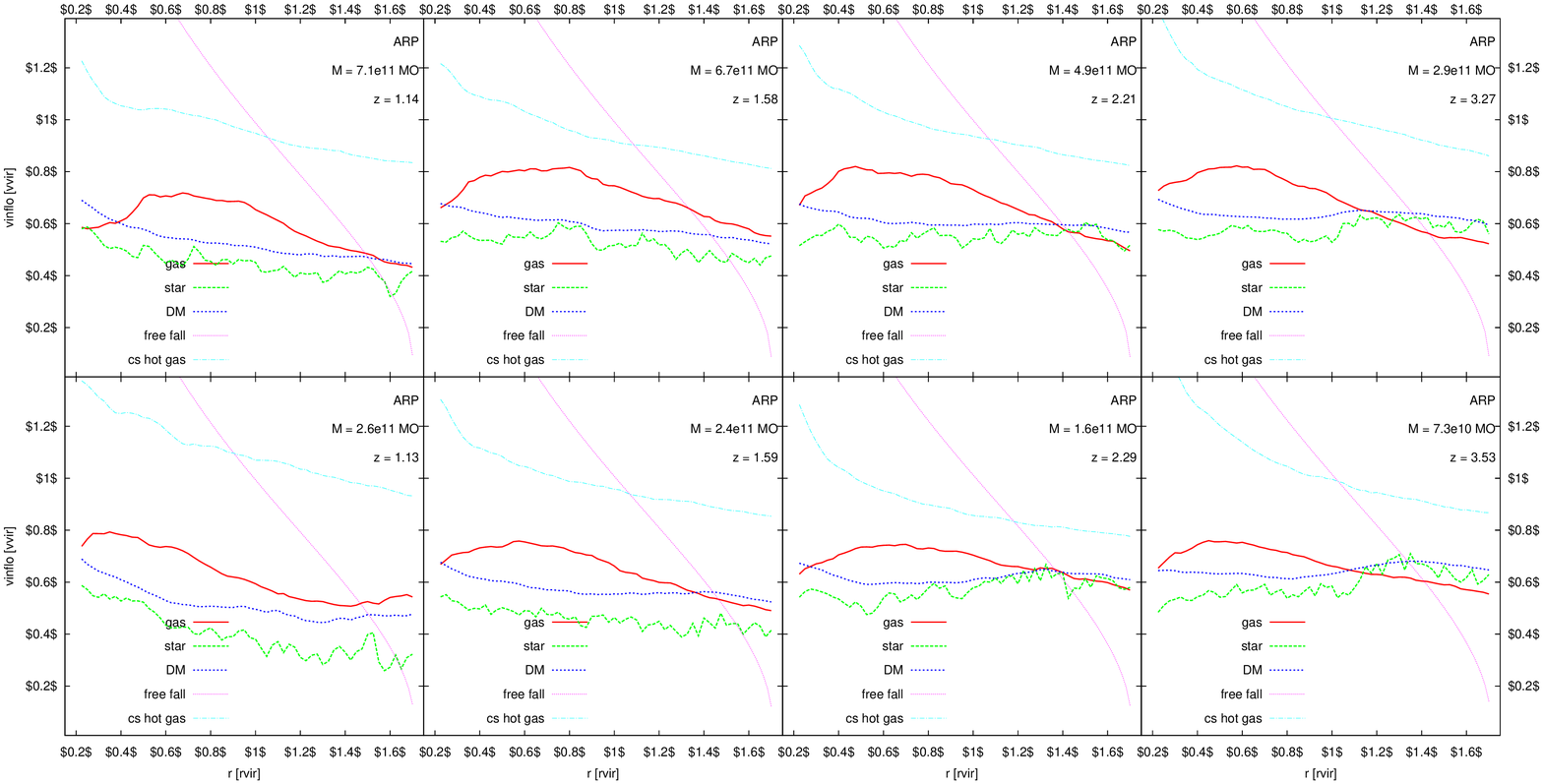}
\end{center}
\caption{Mass weighted averages of the inflow velocities as a function of
radius for the {\C14} simulations. Free-fall profiles are over-plotted in
dotted magenta  and the sound speeds of the hot ambient halo gas in dash-dotted
turquoise. Again the inflow velocities of the gas increase with decreasing
radius in most cases. These increases are again not consistent with free-fall.
In most cases one sees an increase of velocity with decreasing radius down to
$\sim 0.5 \ r_{\rm vir}$ from where the velocity is decreasing again. A shape
like that is typical for low mass haloes at high redshift.}
\label{fig:inflvelvsradARP}
\end{figure*}

For a first order approximation the temperature of the ambient halo gas is
assumed to be equal to the virial temperature of the host halo. It is constant
with position and given by
\begin{equation}
T_{\rm vir} = {G \ M_{\rm vir} \ m_{\rm p} \ \mu_{\rm av} \over 2 \ r_{\rm vir} \
k_{\rm B}}.
\end{equation}
$G$ is the gravitational constant, $M_{\rm vir}$ is the virial mass of the host
halo and $r_{\rm vir}$ its virial radius. These three quantities are related via
\begin{equation}
v_{\rm vir} = \sqrt {G \ M_{\rm vir} \over r_{\rm vir}}.
\label{eqn:vvir}
\end{equation}
After inserting the expressions for $T_{\rm vir}$ and $v_{\rm vir}$ into equation
(\ref{eqn:cs}) we get an expression for the sound speed. Assuming that the
velocity of the gas in the cold streams indeed equals to the sound speed of the
ambient medium, because the shocks which occur as soon as the inflow velocity
reaches the sound speed will fundamentally limit the velocity of the gas one
gets a first order approximation for the infall velocity:
\begin{equation}
v_{\rm inflow} = 0.913 \ v_{\rm vir}
\label{eqn:vin}
\end{equation}
In this approximation the infall velocity is constant with radial distance from
the host halo. We will come back to this statement at the end of section
\ref{sec:inflvel} where we will see that this particular equation tends to be
accurate for very high redshift haloes $(z \gtrsim 4.0)$ and more importantly
that the infall velocity is in general indeed constant with radius.

\begin{figure}
\begin{center}
\psfrag{$0.2$}[B][B][1][0]{0.2}
\psfrag{$0.3$}[B][B][1][0]{0.3}
\psfrag{$0.4$}[B][B][1][0]{0.4}
\psfrag{$0.5$}[B][B][1][0]{0.5}
\psfrag{$0.6$}[B][B][1][0]{0.6}
\psfrag{$0.7$}[B][B][1][0]{0.7}
\psfrag{$0.8$}[B][B][1][0]{0.8}
\psfrag{$0.9$}[B][B][1][0]{0.9}
\psfrag{$1$}[B][B][1][0]{1.0}
\psfrag{$1.2$}[B][B][1][0]{1.2}
\psfrag{$1.4$}[B][B][1][0]{1.4}
\psfrag{$1.6$}[B][B][1][0]{1.6}
\psfrag{r [rvir]}[B][B][1][0]{$r$ [$r_{\rm vir}$]}
\psfrag{vinflo [vvir]}[B][B][1][0]{$v_{\rm inflow}$ [$v_{\rm vir}$]}
\psfrag{MN: 1e12 MO, z = 2.46}[Br][Br][1][0]{\textcolor{blue}{{\MN}:
$10^{12}\,$M$_\odot$, $z = 2.46$}}
\psfrag{CDB: 8.6e11 MO, z = 2.25}[Br][Br][1][0]{\textcolor{red}{{\CD}:
$8.6 \times 10^{11}\,$M$_\odot$, $z = 2.25$}}
\psfrag{ARP: 4.9e11 MO, z = 2.21}[Br][Br][1][0]{\textcolor{green}{{\C14}:
$4.9 \times 10^{11}\,$M$_\odot$, $z = 2.21$}}
\psfrag{gas}[Bl][Bl][1][0]{gas}
\psfrag{stars}[Bl][Bl][1][0]{stars}
\psfrag{DM}[Bl][Bl][1][0]{DM}
\psfrag{free fall}[Br][Br][1][0]{\textcolor{magenta}{free-fall}}
\psfrag{cs hot gas}[Br][Br][1][0]{\textcolor{cyan}{$c_{\rm s}$ hot gas}}
\includegraphics[width=7.34cm]{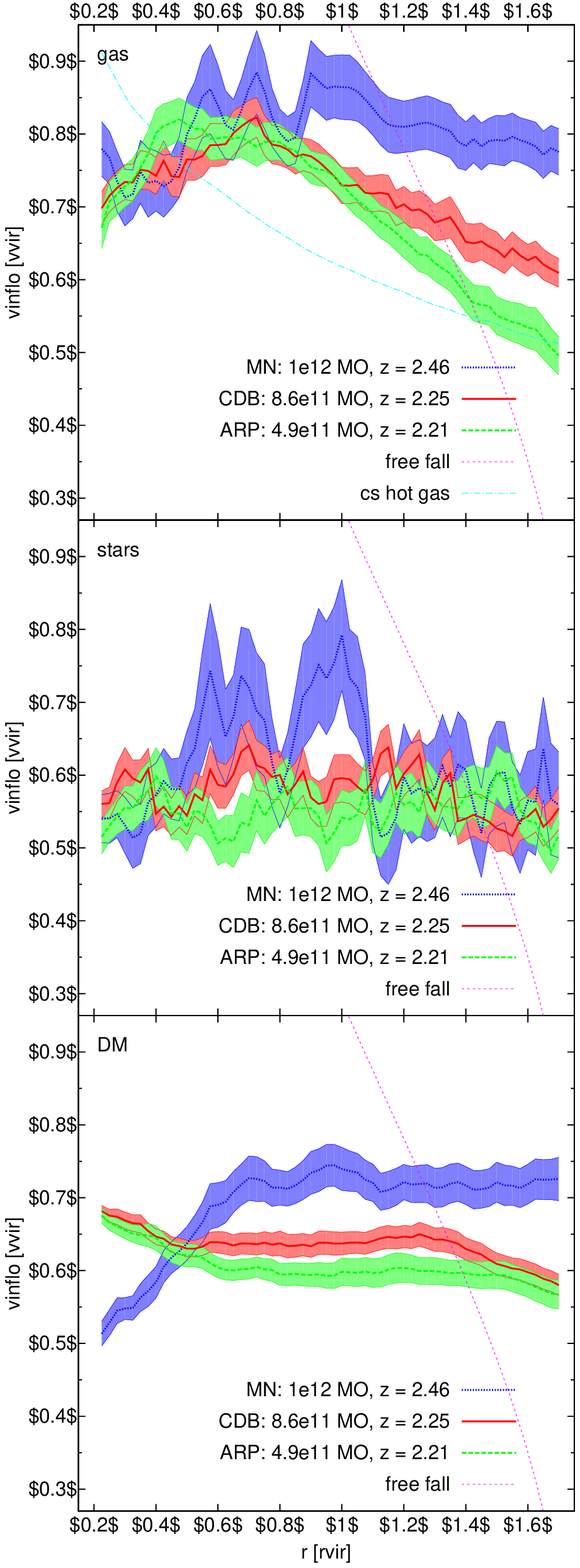}
\end{center}
\caption{Direct comparison between the {\MN} (dotted blue), the {\CD} (solid
red) and the {\C14} (long-dashed green) suite of simulations. The inflow
velocities as a function of radius are shown together with their 1 $\sigma$
standard deviation lines. Over-plotted are a free-fall profile (short-dashed
magenta) and the sound speed of the hot ambient halo gas $c_{\rm s}$
(dash-dotted turquoise). Inflow velocities for {\MN}, {\CD} and {\C14} roughly
agree with each other within the 1 $\sigma$ standard deviation, small offsets
can be accounted for by the slightly different redshifts.}
\label{fig:inflvelvsradcombi}
\end{figure}

\subsubsection{Simulated halo gas temperature}

A more sophisticated measure of the ambient medium's temperature is not to
approximate the gas temperature anymore, but to measure it directly from the
hydrodynamic simulations as a function of position. Since we assume spherical
symmetry it is sufficient to measure it as a function of radial distance from
the host halo centre. When looking at the simulations one has to carefully
discriminate between the gas of the ambient medium which is hot and the gas
flowing into the host as part of the cold streams which is cold. Only the
temperatures of the ambient medium will determine the velocity of the cold
streams. To distinguish between cold streams on the one hand and hot halo gas
on the other hand we adopt the following temperature cut: temperatures are
defined as 'hot' in this sense if they are at least $> 0.2 \ T_{\rm vir}$ or $>
3 \times 10^4$ K whatever is the higher value. The mass weighted average
temperatures of the hot ambient halo gas only are then taken in spherically
symmetric bins at different radii $r$ to get a numerical temperature profile
$T(r)$. Inserting this numerical profile for the temperature into equation
(\ref{eqn:cs}) we get a sound speed profile for each set of simulated galaxies
that are listed in table \ref{tab:bins}. The sound speed profiles are presented
together with the results from the hydrodynamical simulations in section
\ref{sec:inflvel}. For the {\MN} simulations at intermediate to high masses
$(M_{\rm vir} \ge 10^{12}$ M$_\odot)$, the sound speed is increasing with
decreasing redshift almost as sharply as the free-fall profiles. In these
regions of the parameter space it can only serve as a slightly better proxy for
the inflow velocity than free-fall. In all four bins of the {\CD} simulations
the sound speed profiles increase slightly with decreasing radius. In case of
the {\CD} simulations the sound speeds match the simulations very precisely as
we will see later in figure \ref{fig:inflvelvsradCDB}. The profiles increase
slightly with decreasing radius also in the {\C14} simulations. Their shape is
very similar to the actual simulation values but the sound speeds are usually
50\% higher than the simulation values. To summarise: the sound speed of the
hot ambient halo gas is a better proxy for the velocity of the infall than
free-fall. It is more precise at low masses $(M_{\rm vir} \le 10^{12}$ M$_\odot)$
and gives very good results for the {\CD} suite of simulations. However it is
only an incomplete prescription for the inflow velocity.

\section{Inflow velocities}
\label{sec:inflvel}

The analysis of the hydrodynamical simulation starts by compiling averaged
infall velocity profiles for all simulated galaxies listed in table
\ref{tab:bins}. Since we want to look at the velocities of the inflowing
cold streams and not at the average velocity of some net inflow we have to
discriminate between the inflowing material belonging to the cold streams and
the ambient halo material. We experimented with different cuts and found out
that a simple inflow criterion should be sufficient. Since different gas phases
are reported to have very different accretion properties \citep{nelsona} and we
want to concentrate on the kinematics of the cold flows only we decided to
adopt for the gas an additional temperature cut of $T < 3 \times 10^4$ K. So
the velocities are computed for gas by averaging the radial velocity component
of all inflowing cells having temperatures less than $3 \times 10^4$ K in a
given radius bin. The dark matter and stellar velocities are computed for by
averaging the radial velocity component of all inflowing particles in a given
radius bin. Only cells and particles with negative radial velocities and only
cells with temperatures less than $3 \times 10^4$ K are taken into account,
i.e. all motion perpendicular to the radial infall direction as well as all
cells or particles which are moving outwards as well as all cells with
temperatures higher than $3 \times 10^4$ are completely neglected. The
resulting values are averaged over all available galaxies in the respective
mass and redshift bin. Velocities are always quoted dimensionless, namely in
units of the virial velocity $v_{\rm vir}$ which has already been defined in
equation (\ref{eqn:vvir}).

The resulting inflow velocity profiles are shown in figures
\ref{fig:inflvelvsradMN} (\MN), \ref{fig:inflvelvsradCDB} (\CD) and
\ref{fig:inflvelvsradARP} (\C14). Free-fall profiles are over-plotted in dotted
magenta and the sound speeds of the ambient gas is over-plotted in dash-dotted
turquoise. The most striking feature visible is that the velocity is to a very
good approximation constant with radius. It is clearly not free-falling. This
is the case for all three suites of simulations, for all redshifts or masses as
well as for all three constituents: gas, stars or dark matter. The potential
energy of the inflowing gas is not converted into kinetic energy, it has to be
dissipated by another mechanism, most likely by ly$\alpha$-radiation
\citep{mich}. In some of the cases however the velocity of the gas follows the
sound speed profiles of the hot ambient medium fairly well.

Looking at the graphs more carefully the following deviations from constant
velocity infall can be found: {\MN}'s high mass haloes $(M_{\rm vir} \ge 5
\times 10^{12}$ M$_\odot)$ have decreasing inflow velocities with decreasing
radius. In most of the {\CD} and {\C14} galaxies on the other hand an increase
of the velocity can be seen as the material falls further in. This infall
however is far from being consistent with free-fall as the overlain profiles
demonstrate. In all other cases the inflow velocity is indeed constant.

Additionally there is a dependency of the shape of the infall velocity profiles
with mass and redshift: For low mass haloes $(M_{\rm vir} \le 5 \times 10^{12}$
M$_\odot)$ at high redshift $(z \gtrsim 1.7)$ usually there is an increase of
velocity with decreasing radius down to $\sim 0.5 \ r_{\rm vir}$ from where the
velocity is decreasing again. This behaviour originates in the loss of angular
momentum due to strong torques in a tilted extended ring, the ``strong-torque
phase'' of the inflow \citep[][their section 5 and figure 20]{mark}. For very
high mass haloes $(M_{\rm vir} > 2.5 \times 10^{12}$ M$_\odot)$ the inflow
velocity is slightly increasing with increasing radius over the whole radius
range without ever turning down again. In the remaining bins the inflow
velocity is constant or only very mildly increasing with decreasing radius also
over the whole radius range also without ever turning down again.

The velocities of the star particles with negative radial velocities stay
remarkably constant over the whole radius range for all masses and redshifts.
The velocities of the dark matter particles with negative radial velocities
increase with increasing radius for high mass galaxies $(>3 \times 10^{12}$
M$_\odot)$ and decrease with increasing radius for low mass galaxies $(<3\times
10^{12}$ M$_\odot)$. One would naively expect that at least the dark matter
particles inflow velocities would follow free-fall, since those are
collision-less and therefore cannot be slowed down by any baryonic processes.
However there is no meaningful dark matter accretion within $\sim 2 \ r_{\rm
vir}$ as already pointed out by \citet{cuesta} or \citet{wetzel}. Therefore the
dark matter particles are not inflowing within the radius range analysed, they
are just moving in and out again of a certain innermost sphere since the halo
is virialised and most of the dark matter particles are orbiting on
non-circular orbits.

\begin{figure*}
\begin{center}
\psfrag{z + 1}[B][B][1][0]{$z + 1$}
\psfrag{vinflow [vvir]}[B][B][1][0]{$v_{\rm inflow}$ [$v_{\rm vir}$]}
\psfrag{Mvir [MO]}[B][B][1][0]{$M_{\rm vir}$ [M$_\odot$]}
\psfrag{MN}[Br][Br][1][0]{\MN}
\psfrag{CDB}[Br][Br][1][0]{\CD}
\psfrag{ARP}[Br][Br][1][0]{\C14}
\psfrag{$0.5$}[B][B][1][0] {$0.5$}
\psfrag{$0.6$}[B][B][1][0] {$0.6$}
\psfrag{$0.7$}[B][B][1][0] {$0.7$}
\psfrag{$0.8$}[B][B][1][0] {$0.8$}
\psfrag{$0.9$}[B][B][1][0] {$0.9$}
\psfrag{$2.5$}[B][B][1][0] {$2.5$}
\psfrag{$3$}[B][B][1][0] {$3$}
\psfrag{$3.5$}[B][B][1][0] {$3.5$}
\psfrag{$4$}[B][B][1][0] {$4$}
\psfrag{$4.5$}[B][B][1][0] {$4.5$}
\psfrag{$5$}[B][B][1][0] {$5$}
\psfrag{$1e11$}[B][B][1][0]{$10^{11}$}
\psfrag{$1e12$}[B][B][1][0]{$10^{12}$}
\psfrag{$1e13$}[B][B][1][0]{$10^{13}$}
\psfrag{gas}[Br][Br][1][0]{gas}
\psfrag{scaled to 1e12}[Br][Br][1][0]{scaled to $M_{\rm vir} = 10^{12}$ M$_\odot$}
\psfrag{scaled to z = 2.46}[Br][Br][1][0]{scaled to $z = 2.46$}
\includegraphics[width=17.73cm]{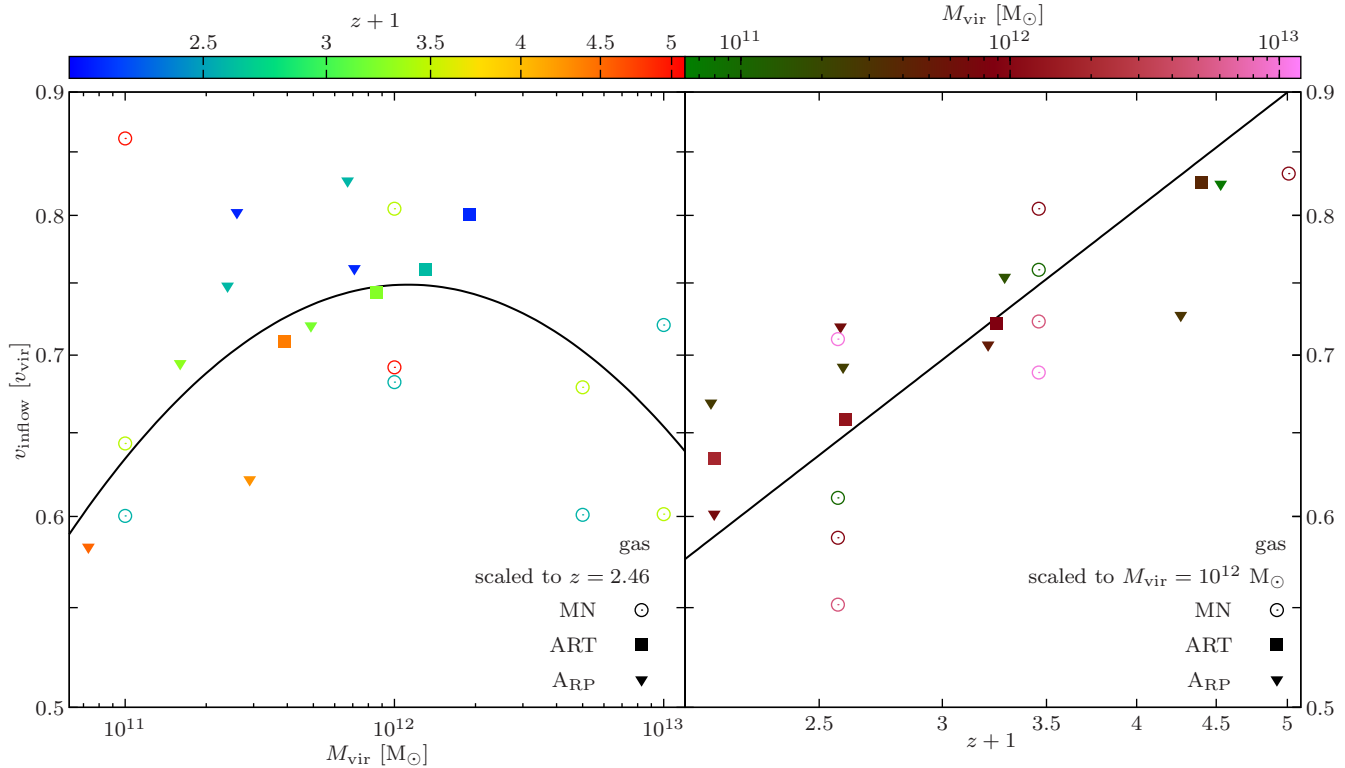}
\end{center}
\caption{Average inflow velocity $v_{\rm inflow}$ in units of the virial velocity
as a function of virial mass $M_{\rm vir}$ ({\it left}) or as a function of
redshift $z + 1$ ({\it right}). There is a strong square root power law
relationship between the inflow velocity and the redshift ({\it right panel}).
The inflow velocity also follows a ``parabola-like'' relation with respect to
the host halo mass on the other hand that peaks roughly at $M_{\rm vir} =
10^{12}$ M$_\odot$ ({\it left panel}). The data points within each panel were
scaled with the help of the scaling relation in the respective other panel to
the values indicated. Both scaling relations were found together in an
iterative process (see text). The colour bar axes indicate the values the data
points used to have before rescaling.}
\label{fig:vvirinflowfit}
\end{figure*}

The finding of constant velocity inflow is in contrast to earlier works
\citep{faucher, joki}. These authors report evidence for free-fall behaviour of
the cold streams. A possible explanation for the discrepancies with
\citet{faucher} might be, that they deployed \textsc{sph} simulations which
used to have inaccuracies modelling dissipative processes \citep{suckiness}.
Such processes are taking place frequently at the boundary layers of cold
streams. Also \citet{nelsona} found differences in the accretion properties for
different hydro schemes. \citet{joki} on the other hand used \textsc{amr}
techniques, like us. Their sample of galaxies however was small (only three
galaxies in total) and they report that not all of their galaxies showed signs
of free-fall behaviour. Constant velocity inflow on the other hand is in
agreement with the results from \citet{teklu}. She finds constant velocity
profiles at around 170\,km\,s$^{-1}$ for three galaxies with $M_{\rm vir} = 1.5
\times 10^{12}$\,M$_\odot$ at $z = 2.33$ in her simulations. Roughly constant
velocity inflow is also in agreement with the results of \citet{freke}, their
absolute velocities of 150 km/s $(\sim 0.75 \ v_{\rm vir})$ is remarkably close
to the values we found $(\sim 0.7 \ v_{\rm vir})$. Also \citet{nelsonc} find
roughly constant inflow velocities, c.f their figure 7. Strictly non
free-falling inflow is also in agreement with the results from \citet[][their
figures 4 and 6]{wetzel} who see a decrease in velocity with decreasing radius
(roughly a decrease of 65\% between 1 $r_{\rm vir}$ and the centre of the host).
Unlike us they did not exclude outflowing cells and particles which explains
their velocities that are varying with radius as well as the on average lower
values of their speeds.

\begin{figure}
\begin{center}
\includegraphics[width=8.45cm]{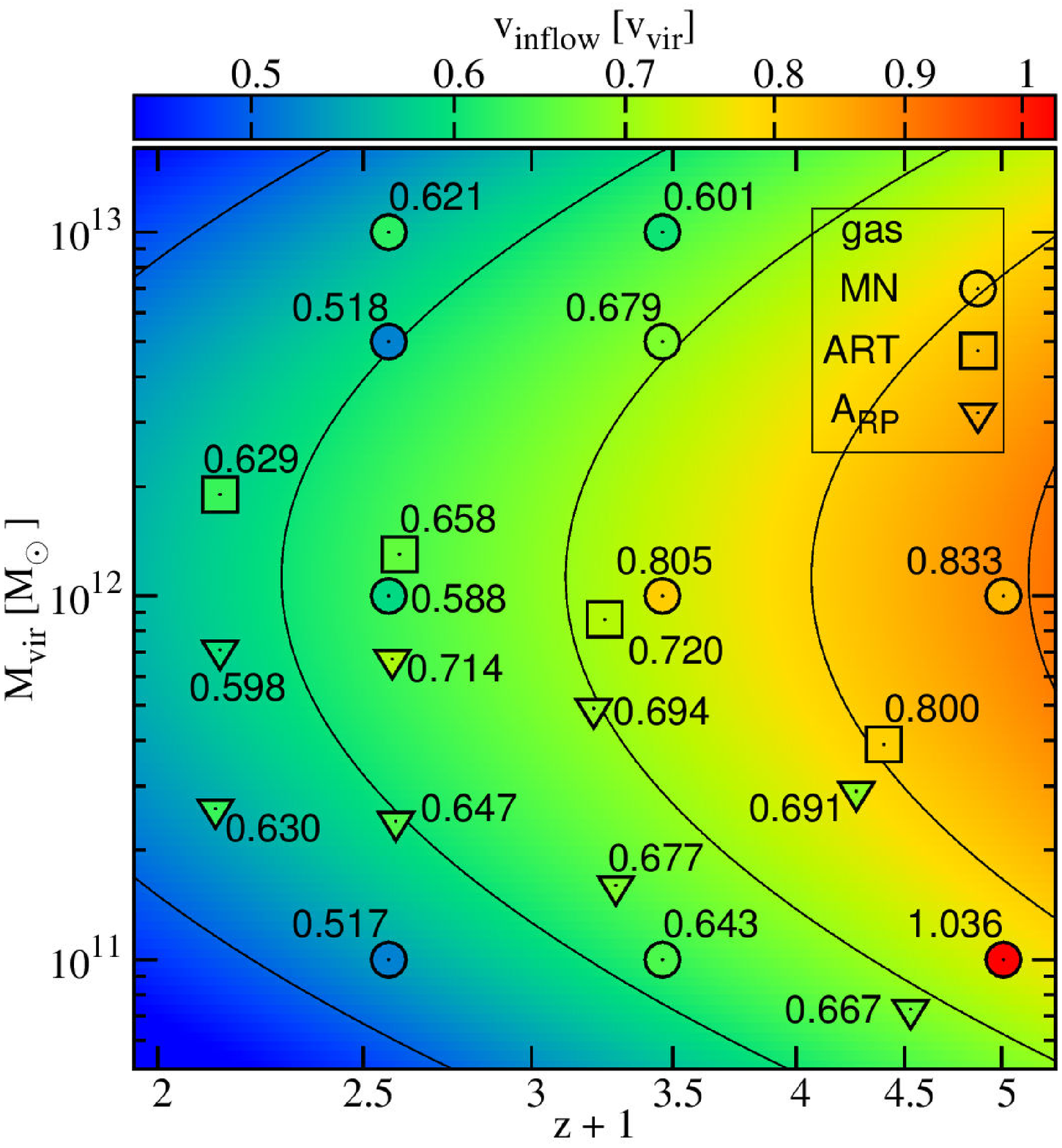}
\end{center}
\caption{Average inflow velocity of the gas in units of the virial velocity as
a function of halo mass and redshift. Shown as background colour is the fitting
function (equation \ref{eqn:vvirinflow}) with contour lines at 0.5, 0.6, 0.7,
0.8 and 0.9 $v_{\rm vir}$. The actual values measured from the simulations are
given with colour coding within the open symbols as well as by the attached
labels. {\CD} is marked by squares, {\MN} by circles and {\C14} by triangles.}
\label{fig:vinflowfit3D}
\end{figure}

To guarantee the consistency of the {\MN}, the {\CD} and the {\C14} simulations 
it is necessary to compare them directly. In figure \ref{fig:inflvelvsradcombi}
galaxies at $z \sim 2.3$ with $M_{\rm vir} \sim 8 \times 10^{11}$ M$_\odot$ are
shown for the three suites of simulations. The lines for stars and dark matter
are very similar for all three: the velocities are roughly constant over the
whole radius range at comparable values. However the gas curves are slightly
different: the {\MN} one is by and large constant, whereas the {\CD} as well as
the {\C14} one increase slightly with decreasing radius. The overall increase
is roughly 35 \% over the whole radius range $(1.7 - 0.2 \ r_{\rm vir})$. These
increases, as mentioned earlier, are not consistent with free-fall. There is an
offset in the values: the {\MN} velocities are $\sim 30\%$ higher over most of
the radius range. The offset can be accounted for by the offsets, which the
host haloes' masses and redshifts have.

\citet{nelsonb} found that feedback affects gas accretion. Our three suites of
simulations use varying strengths of feedback. As we can see from figure
\ref{fig:inflvelvsradcombi} varying strengths of feedback affect the inflow
velocities only very mildly. However as we will discuss in our forthcoming
companion paper \citep{mich5} varying strengths of feedback do affect the total
amount of inflow.

After the consistency of the infall velocities in all three suites of
simulations has been established one can now look at possible trends of the
infall velocities with mass and redshift. We want to quantify the qualitative
statements made earlier. For simplicity a constant infall velocity with radius
is assumed. Its actual value is determined by a constant line fit through the
velocity curve of the respective mass-redshift bin.

The average inflow velocities $v_{\rm inflow}$ of the gas are shown as a function
of halo mass $M_{\rm vir}$ and redshift $z+1$ in figure \ref{fig:vvirinflowfit}.
Plotted are in both panels all available bins of galaxies from {\MN}, {\CD} and
{\C14} as listed in table \ref{tab:bins}. In the left panel the $v_{\rm inflow}$
values are plotted against the halo mass $(M_{\rm vir})$ and in the right panel
against redshift $z + 1$. One can easily identify that the $v_{\rm inflow}$
values in the right panel follow a linear relation with respect to $z + 1$ on
this log-log graph, which is a power law. Since in this case the exponent is
remarkably close to a half we decided in order to reduce the number of free
parameters of our fitting function to choose a square root power law as the
fitting function for the $v_{\rm inflow}$ values with respect to $z + 1$. The
best fit square root power law is over-plotted by the solid black line. The
$v_{\rm inflow}$ values in the left panel however seem to follow a
``parabola-like'' function on this log-log graph, which is the log-normal
distribution function. We choose this function because the parabola is the
simplest function with a non-constant derivative, i.e. the simplest function
which rises first and declines later. Since we are working in log-log space we
had to go with its log-log analogue, the log-normal distribution function. The
function peaks at $M_{\rm vir} = 10^{12}$ M$_\odot$. For the sake of simplicity we
will refer to this kind of function as ``parabola-like'' throughout the paper.

The $v_{\rm inflow}$ values of most data points in both panels had to be rescaled
in the following way: the $v_{\rm inflow}$ values of the data points within the
left panel that were not at $z = 2.46$ were scaled to this value according to
the scaling relation presented in the right panel. Likewise the $v_{\rm inflow}$
values of the data points within the right panel whose underlying galaxy bins
were not at $M_{\rm vir} = 10^{12}$ M$_\odot$ had to be rescaled to that value
according to the scaling relation presented in the left panel. The colour bar
axes in both panels indicate the values the galaxy bins used to have before
rescaling.

Both scaling relations were found in the following iterative process: first a
square root power law was fit through the yet unrescaled data points in the
right panel. This relation was then been used to rescale the data points in the
left panel. Subsequently a log-normal distribution function was fit through the
(scaled) data points in the left panel. Afterwards the resulting log-normal
distribution function was used to rescale the data points in the right panel.
These steps have been repeated until convergence was reached (i.e. until the
resulting scaling relations did not change anymore). Convoluting the square
root power law with the log-normal distribution function leads to the following
equation which serves as a model describing the behaviour of the cold stream
infall velocity as a function of redshift and host halo mass:
\begin{eqnarray}
{v_{\rm inflow}(M_{\rm vir},z) \over v_{\rm vir}(M_{\rm vir},z)} & = & {A \
\sqrt{z + 1} \over \sigma \ (M_{\rm vir} / {\rm M}_\odot)} \nonumber \\
& \times & \exp \left\{-{\left[\ln (M_{\rm vir} / {\rm M}_\odot) - \mu \right]^2
\over 2 \ \sigma^2}\right\}
\label{eqn:vvirinflow}
\end{eqnarray}
Best fit parameter are $A = (1.25 \pm 0.13) \times 10^{16}$, $\mu = 45.4 \pm
4.8$ and $\sigma =4.20 \pm 0.53$. A visualisation of this 3D functional fit is
shown in figure \ref{fig:vinflowfit3D}, where the values that are actually
measured from the simulations are given by the colour within the open symbols
as well as by the attached labels. The {\CD} results are represented by
squares, the {\C14} results by triangles and the {\MN} results by circles.
Shown as background colour is the fitting function (equation
\ref{eqn:vvirinflow}) with contour lines at 0.5, 0.6, 0.7, 0.8 and 0.9
$v_{\rm vir}$.

We are aware of the fact that the choice of a ``parabola-like'' function
instead of another power law as a scaling relation for the data points in the
left hand side panel of figure \ref{fig:vvirinflowfit} hinges only on the four
very high mass $(M \ge 5 \times 10^{12}$ M$_\odot)$ data points of the {\MN}
simulation which is the one having the lowest resolution. We decided to trust
these data points anyway since they were at the high mass end of our collection
of galaxies were resolution issues play only a minor role. Unfortunately they
were no {\CD} or {\C14} galaxies available in this mass range. On the other
hand we decided to neglect the {\MN} data point at $M_{\rm vir} = 10^{11}$
M$_\odot$ and $z = 4.01$ having a very high inflow velocity as an outlier since
it has a very low mass, so it might be more prone to resolution effects.

We are also aware of the fact that our choice of functional forms for equation
(\ref{eqn:vvirinflow}) as well as figures \ref{fig:vvirinflowfit} and
\ref{fig:vinflowfit3D} the square root power law on the one hand and the
log-normal distribution on the other hand might seem to be very specific and
arbitrary but those are the log-log analogues of the simplest functional forms 
having constant and non-constant derivatives, the straight line and the
parabola. It is surely possible to find other functional forms which would lead
to a different equation (\ref{eqn:vvirinflow}), but due to the constraining
power of the data points those alternative functional forms should ultimately
arrive at very similar predictions (i.e. the almost same appearance of figure
\ref{fig:vinflowfit3D}) but they will have more free parameters.

\section{Conclusions}
\label{sec:conc}
In this paper we looked at the radial inflow velocities of accretion along
streams from the cosmic web into massive galaxies at high redshifts using three
sets of \textsc{amr} hydro-cosmological simulations. We calculated free-fall
profiles as well as sound speeds of the hot ambient medium via two different
methods and we found the following:

\begin{itemize}

\item The velocity profiles for the cold streams in the simulations are very
different from free-fall.

\item Not even the infall velocities of the dark matter particles follow
free-fall, since there is no meaningful dark matter accretion within $\sim 2 \
r_{\rm vir}$ of the central halo \citep{cuesta, wetzel}. Therefore the dark
matter particles are not inflowing they are rather moving back and forth.

\item The sound speed of the ambient medium is a better proxy for the behaviour
of the gas inflow velocity, at least at high redshifts. We present a neat
scaling relation for the inflow velocity based on this as a first order
approximation $(v_{\rm inflow} = 0.9 \ v_{\rm vir})$.

\item The velocity profiles for the cold streams are by and large constant with
radius or only very slowly increasing with decreasing radius.

\item There is a dependence of the shape of the gas infall velocity profiles
with mass and redshift: for low mass haloes $(M_{\rm vir} \le 5 \times 10^{12}$
M$_\odot)$ at higher redshift $(z \gtrsim 1.7)$ one usually sees an increase of
velocity with decreasing radius down to $\sim 0.5 \ r_{\rm vir}$ from where the
velocity is decreasing again. This originates in the loss of angular momentum
during the ``strong-torque phase'' \citep{mark}.

\item For very high mass haloes $(M_{\rm vir} > 2.5 \times 10^{12}$ M$_\odot)$ the
inflow velocity is slightly increasing with increasing radius over the whole
radius range. In all other cases the inflow velocity is constant or very mildly
increasing with decreasing radius also over the whole radius range.

\item This constant infall velocity has in units of the virial velocity as a
function of radius a ``parabola-like'' dependency on the host halo mass that
peaks at $M_{\rm vir} = 10^{12}$ M$_\odot$ and it also follows a square root power
law relations with respect to the redshift $(v_{\rm inflow} \propto \sqrt{z + 1} \
v_{\rm vir})$.

\item Since the magnitude of the mass accretion rate is also constant with
radius one can follow that cold and hot gas do not mix as the cold streams fall
in.

\end{itemize}

A potential limitation of our simulations could arise from the artificial
pressure floor used in the simulations to ensure that the Jeans mass is always
fully resolved. It might affect the temperature and density of the very dense
and cold parts of the streams, with implications on the computed inflow
velocities. Also the interaction between outflows and inflows is yet to be
studied in simulations with strong feedback. Indeed preliminary results of
\citet{house} show that there is no strong effect of strong feedback on to
inflows. But \textsc{amr} codes are still the best available tool for
recovering the stream properties. With their 17 pc resolution, and with proper
cooling below $10^4$K, these simulations provide the most reliable description
of the cold streams that are available so far.

We conclude that the velocity profile of the gas flowing into a galaxy's halo
in the form of cold streams is, contrary to what might be expected, roughly
constant with radius instead of free-falling. The potential energy of the gas
which is lost on its way is not converted into kinetic energy but must be
dissipated by other mechanisms, such as Ly$\alpha$ radiation \citep{mich}.

\section*{Acknowledgements}
Tobias Goerdt is a Lise Meitner fellow. We thank Romain Teyssier for his
kindness in sharing simulation data with us. We acknowledge stimulating
discussions with Yuval Birnboim, Nicolas Bouch{\'e} and Oliver Czoske. The
simulations were performed at the astro cluster at the Hebrew University of
Jerusalem, at the National Energy Research Scientific Computing Centre,
Lawrence Berkeley National Laboratory and at NASA Advanced Supercomputing at
NASA Ames Research Center. Parts of the computational calculations were done at
the Vienna Scientific Cluster under project number 70522. The authors would
like to thank Jorge S{\'a}nchez Almeida and the Instituto de Astrof{\'i}sica de
Canarias for their hospitality, where parts of this work were carried out. This
work was supported by FWF project number M 1590-N27.

\bibliographystyle{mn2e}
\bibliography{infvel05.bbl}

\begin{thebibliography}{}

\bibitem[\protect\citeauthoryear{Agertz et al.}{2007}]{suckiness}
Agertz O. et al, 2007, MNRAS, 380, 963

\bibitem[\protect\citeauthoryear{Agertz, Teyssier \& Moore}{2009}]{oscara}
Agertz O, Teyssier R, Moore B, 2009, MNRAS, 397L, 64

\bibitem[\protect\citeauthoryear{Agertz, Teyssier \& Moore}{2011}]{oscarb}
Agertz O, Teyssier R, Moore B, 2011, MNRAS, 410, 1391

\bibitem[\protect\citeauthoryear{Agertz et al.}{2013}]{oscarc}
Agertz O, Kravtsov A. V, Leitner S. N, Gnedin N. Y, 2013, ApJ, 770, 25

\bibitem[\protect\citeauthoryear{Basu-Zych \& Scharf}{2004}]{basu}
Basu-Zych A, Scharf C, 2004, ApJ, 615, 85

\bibitem[\protect\citeauthoryear{Birnboim \& Dekel}{2003}]{bd03}
Birnboim Y, Dekel A, 2003, MNRAS, 345, 349

\bibitem[\protect\citeauthoryear{Bouch{\'e} et al.}{2010}]{bouche}
Bouch{\'e} N. et al, 2010, ApJ, 718, 1001

\bibitem[\protect\citeauthoryear{Bouch{\'e} et al.}{2013}]{bouche2}
Bouch{\'e} N, Murphy M. T, Kacprzak G. G, P{\'e}roux C, Contini T, Martin C. L,
Dessauges-Zavadsky M, 2013, Science, 341, 50

\bibitem[\protect\citeauthoryear{Cacciato, Dekel \& Genel}{2012}]{marcello}
Cacciato M, Dekel A, Genel S, 2012, MNRAS, 421, 818

\bibitem[\protect\citeauthoryear{Ceverino \& Klypin}{2009}]{cak}
Ceverino D, Klypin A. A, 2009, ApJ, 695, 292

\bibitem[\protect\citeauthoryear{Ceverino, Dekel \& Bournaud}{2010}]{cd}
Ceverino D, Dekel A, Bournaud F, 2010, MNRAS, 404, 2151

\bibitem[\protect\citeauthoryear{Ceverino et al.}{2012}]{c12}
Ceverino D, Dekel A, Mandelker N, Bournaud F, Burkert A, Genzel R, Primack J,
2012, MNRAS, 420, 3490

\bibitem[\protect\citeauthoryear{Ceverino et al.}{2014}]{cip}
Ceverino D, Klypin A, Klimek E. S, Trujillo-Gomez S, Churchill C. W, Primack J,
Dekel A, 2014, MNRAS, 442, 1545

\bibitem[\protect\citeauthoryear{Ceverino et al.}{2015}]{c14b}
Ceverino D, Dekel A, Tweed D, Primack J, 2015, MNRAS, 447, 3291

\bibitem[\protect\citeauthoryear{Ceverino-Rodriguez}{2008}]{ceverino}
Ceverino-Rodriguez D, 2008, Ph.D.~Thesis, New Mexico State University

\bibitem[\protect\citeauthoryear{Chapman et al.}{2001}]{chapman}
Chapman S. C, Lewis G. F, Scott D, Richards E, Borys C, Steidel C. C,
Adelberger K. L, Shapley A. E, ApJ, 548, 17

\bibitem[\protect\citeauthoryear{Cuesta et al.}{2008}]{cuesta}
Cuesta A. J, Prada F, Klypin A, Moles M, 2008, MNRAS, 389, 385

\bibitem[\protect\citeauthoryear{Daddi et al.}{2010}]{daddi}
Daddi E. et al, 2010, ApJ, 713, 686

\bibitem[\protect\citeauthoryear{Dalla Vecchia \& Schaye}{2008}]{dvs}  
Dalla Vecchia C, Schaye J, 2008, MNRAS, 387, 1431

\bibitem[\protect\citeauthoryear{Danovich et al.}{2015}]{mark}
Danovich M, Dekel A, Hahn O, Ceverino D, Primack J, 2015, MNRAS, 449, 2087

\bibitem[\protect\citeauthoryear{Dekel \& Birnboim}{2006}]{db06}
Dekel A, Birnboin Y, 2006, MNRAS, 368, 2

\bibitem[\protect\citeauthoryear{Dekel et al.}{2009a}]{DekelA_09a}
Dekel A. et al, 2009a, Nature, 457, 451

\bibitem[\protect\citeauthoryear{Dekel, Sari \& Ceverino}{2009b}]{DekelA_09b}
Dekel A, Sari R, Ceverino D, 2009b, ApJ, 703, 785

\bibitem[\protect\citeauthoryear{Dekel et al.}{2013}]{dekel13}
Dekel A, Zolotov A, Tweed D, Cacciato M, Ceverino D, Primack J. R, 2013,
MNRAS, 435, 999

\bibitem[\protect\citeauthoryear{Dubois \& Teyssier}{2008}]{dubois}
Dubois Y, Teyssier R, 2008, A\&A, 477, 79

\bibitem[\protect\citeauthoryear{Fardal et al.}{2001}]{fardal}
Fardal M. A, Katz N, Gardner J. P, Hernquist L, Weinberg D. H, Dav\'{e} R, 2001,
ApJ, 562, 605

\bibitem[\protect\citeauthoryear{Faucher-Giguere et al.}{2010}]{faucher}
Faucher-Giguere C. A, Kere{\v s} D, Dijkstra M, Hernquist L, Zaldarriaga M,
2010, ApJ, 725, 633

\bibitem[\protect\citeauthoryear{Ferland et al.}{1998}]{ferland}
Ferland G. J, Korista K. T, Verner D. A, Ferguson J. W, Kingdon J. B, Verner E.
M, 1998, PASP, 110, 761

\bibitem[\protect\citeauthoryear{Fumagalli et al.}{2011}]{fumagalli}
Fumagalli M, Prochaska J. X, Kasen D, Dekel A, Ceverino D, Primack J. R, 2011,
MNRAS, 418, 1796

\bibitem[\protect\citeauthoryear{Geach et al.}{2005}]{geacha}
Geach J. E. et al, 2005, MNRAS, 363, 1398

\bibitem[\protect\citeauthoryear{Geach et al.}{2007}]{geachb}
Geach J. E, Smail I, Chapman S. C, Alexander D. M, Blain A. W, Stott J. P,
Ivison R, 2007, ApJ, 655, 9

\bibitem[\protect\citeauthoryear{Genel, Dekel \& Cacciato}{2012}]{genel12}
Genel S, Dekel A, Cacciato M, 2012, MNRAS, 425, 788

\bibitem[\protect\citeauthoryear{Genzel et al.}{2010}]{genzel10}
Genzel R. et al, 2010, MNRAS, 407, 2091

\bibitem[\protect\citeauthoryear{Genzel et al.}{2011}]{genzel11}
Genzel R. et al, 2011, ApJ, 733, 101

\bibitem[\protect\citeauthoryear{Goerdt et al.}{2010}]{mich}
Goerdt T, Dekel A, Sternberg A, Ceverino D, Teyssier R, Primack J. R,
2010, MNRAS, 407, 613

\bibitem[\protect\citeauthoryear{Goerdt et al.}{2012}]{mich2}
Goerdt T, Dekel A, Sternberg A, Gnat O, Ceverino D, 2012, MNRAS, 424, 2292

\bibitem[\protect\citeauthoryear{Goerdt, Burkert \& Ceverino}{2013}]{mich3}
Goerdt T, Burkert A, Ceverino D, 2013, arXiv:1307.2102

\bibitem[\protect\citeauthoryear{Goerdt et al.}{2015}]{mich5}
Goerdt T, Ceverino D, Dekel A, Teyssier R, 2015, arXiv:1505.01486

\bibitem[\protect\citeauthoryear{Grevesse \& Sauval}{1998}]{grevesse}
Grevesse N, Sauval A. J, 1998, SSRv, 85, 161

\bibitem[\protect\citeauthoryear{Haardt \& Madau}{1996}]{haardt}
Haardt F, Madau P, 1996, ApJ, 461, 20

\bibitem[\protect\citeauthoryear{House \& Dekel}{in preparation}]{house}
House H, Dekel A, in preparation

\bibitem[\protect\citeauthoryear{Ibata et al.}{2013}]{ibata}
Ibata R. A. et al, Nature, 2013, 493, 62

\bibitem[\protect\citeauthoryear{Katz, Hernquist \& Weinberg}{1992}]{katz}
Katz N, Hernquist L, Weinberg D. H, 1992, ApJ, 399, 109

\bibitem[\protect\citeauthoryear{Kennicutt}{1998}]{kennicutt}
Kennicutt R. C, 1998, ApJ, 498, 541

\bibitem[\protect\citeauthoryear{Kere{\v s} et al.}{2005}]{keresa}
Kere{\v s} D, Katz N, Weinberg D. H, Dav{\'e} R, 2005, MNRAS, 363, 2

\bibitem[\protect\citeauthoryear{Kere{\v s} et al.}{2009}]{keresb}
Kere{\v s} D, Katz N, Fardal M, Dav{\'e} R, Weinberg D. 2009, MNRAS, 395, 160

\bibitem[\protect\citeauthoryear{Kravtsov, Klypin \& Khokhlov}{1997}]{kkk}
Kravtsov A. V, Klypin A. A, Khokhlov A. M, 1997, ApJS, 111, 73

\bibitem[\protect\citeauthoryear{Kravtsov}{2003}]{andrey}
Kravtsov A. V, 2003, ApJ, 590, 1

\bibitem[\protect\citeauthoryear{Krumholz \& Burkert}{2010}]{andi}
Krumholz M, Burkert A, 2010, ApJ, 724, 895

\bibitem[\protect\citeauthoryear{Mandelker et al.}{2014}]{mandelker}
Mandelker N, Dekel A, Ceverino D, Tweed D, Moody C. E, Primack J, 2014, MNRAS,
443, 3675

\bibitem[\protect\citeauthoryear{Miller \& Scalo}{1979}]{miller}
Miller G. E, Scalo J. M, 1979, ApJS, 41, 513

\bibitem[\protect\citeauthoryear{Nelson et al.}{2013}]{nelsona}
Nelson D, Vogelsberger M, Genel S, Sijacki D, Kere{\v s} D, Springel V,
Hernquist L, 2013, MNRAS, 429, 3353

\bibitem[\protect\citeauthoryear{Nelson et al.}{2015a}]{nelsonb}
Nelson D, Genel S, Vogelsberger M, Springel V, Sijacki D, Torrey P, Hernquist
L, 2015a, MNRAS, 448, 59

\bibitem[\protect\citeauthoryear{Nelson et al.}{2015b}]{nelsonc}
Nelson D, Genel S, Pillepich A, Vogelsberger M, Springel V, Hernquist L, 2015b,
arXiv:1503.02665

\bibitem[\protect\citeauthoryear{Ocvirk, Pichon \& Teyssier}{2008}]{ocvirk}
Ocvirk P, Pichon C, Teyssier R, 2008, MNRAS, 390, 1326

\bibitem[\protect\citeauthoryear{Prescott et al.}{2015}]{prescott}
Prescott M. K. M, Momcheva I, Brammer G. B, Fynbo J. P. U, M{\o}ller P, 2015,
ApJ, 802, 32

\bibitem[\protect\citeauthoryear{Robertson \& Kravtsov}{2008}]{rk}
Robertson B. E, Kravtsov A. V, 2008, ApJ, 680, 1083

\bibitem[\protect\citeauthoryear{Rosdahl \& Blaizot}{2012}]{joki}
Rosdahl J, Blaizot J, 2012, MNRAS, 423, 344

\bibitem[\protect\citeauthoryear{Scarlata et al.}{2009}]{scarlata}
Scarlata C. et al, 2009, ApJ, 706, 1241

\bibitem[\protect\citeauthoryear{Schaye \& Dalla Vecchia}{2008}]{joop}
Schaye J, Dalla Vecchia C, 2008, MNRAS, 383, 1210

\bibitem[\protect\citeauthoryear{Springel \& Hernquist}{2003}]{sh}
Springel V, Hernquist L, 2003, MNRAS, 339, 289

\bibitem[\protect\citeauthoryear{Steidel et al.}{2011}]{steidel}
Steidel C. C, Bogosavljevic M, Shapley A. E, Kollmeier J. A, Reddy N. A, Erb
D. K, Pettini M, 2011, ApJ, 736, 160

\bibitem[\protect\citeauthoryear{Teklu}{2012}]{teklu}
Teklu A, 2012, Bachelor Thesis, LM University Munich

\bibitem[\protect\citeauthoryear{Teyssier}{2002}]{teyssier}
Teyssier R, 2002, A\&A, 385, 337

\bibitem[\protect\citeauthoryear{Truelove et al.}{1997}]{truelove}
Truelove J. K, Klein R. I, McKee C. F, Holliman J. H, Howell L. H, Greenough J.
A, 1997, ApJ, 489, 179

\bibitem[\protect\citeauthoryear{van de Voort \& Schaye}{2012}]{freke}
van de Voort F, Schaye J, 2012, MNRAS, 423, 2991

\bibitem[\protect\citeauthoryear{Wetzel \& Nagai}{2014}]{wetzel}
Wetzel A. R, Nagai D, 2014, arXiv:1412.0662

\bibitem[\protect\citeauthoryear{Woosley \& Weaver}{1995}]{woosley}
Woosley S. E, Weaver T. A, 1995, ApJS, 101, 181

\bibitem[\protect\citeauthoryear{Yepes et al.}{1997}]{yepes}
Yepes G, Kates R, Khokhlov A, Klypin A, 1997, MNRAS, 284, 235

\end{thebibliography}

\label{lastpage}
\end{document}